\documentclass[11pt]{article}
\usepackage{latexsym} 
\usepackage{amsfonts}
\usepackage{amsmath}
\usepackage{amsthm}
\usepackage{amssymb}
\renewcommand{\theequation}{\thesection.\arabic{equation}}
\newcounter{subequation}[equation]
\makeatletter

\expandafter\let\expandafter\reset@font\csname reset@font\endcsname

\def\subeqnarray{\arraycolsep1pt
	\def\@eqnnum\stepcounter##1{\stepcounter{subequation}%
		{\reset@font\rm(\theequation\alph{subequation})}}
	\jot5mm     \eqnarray}

\makeatother

\def\be{\begin{equation}}
\def\ee{\end{equation}}

\def\lb{\label}

\def\bea{\begin{eqnarray}}
\def\eea{\end{eqnarray}}
\def\ba{\begin{array}}
	\def\ea{\end{array}}
\def\dd{\partial}

\def\half{\frac{1}{2}}

\def\one#1{#1^{\raise5pt\hbox{$\scriptstyle\!\!\!\!1$}}\,{}}
\def\two#1{#1^{\raise5pt\hbox{$\scriptstyle\!\!\!\!2$}}\,{}}

\def\tilde{\widetilde}
\def\II{\hbox{{1}\kern-.25em\hbox{l}}}
\def\a{\alpha}
\def\b{\beta}
\def\c{\gamma}
\def\d{\delta}
\def\e{\varepsilon}
\def\qed{\rule{5pt}{5pt}}
\makeatletter
\def\binrel@#1{\begingroup
	\setboxz@h{\thinmuskip0mu
		\medmuskip\m@ne mu\thickmuskip\@ne mu
		\setbox\tw@\hbox{$#1\m@th$}\kern-\wd\tw@
		${}#1{}\m@th$}%
	\edef\@tempa{\endgroup\let\noexpand\binrel@@
		\ifdim\wdz@<\z@ \mathbin
		\else\ifdim\wdz@>\z@ \mathrel
		\else \relax\fi\fi}%
	\@tempa
}
\let\binrel@@\relax
\def\overset#1#2{\binrel@{#2}%
	\binrel@@{\mathop{\kern\z@#2}\limits^{#1}}}
\def\underset#1#2{\binrel@{#2}%
	\binrel@@{\mathop{\kern\z@#2}\limits_{#1}}}
\makeatother
\newfont{\bbd}{msbm10 scaled\magstep1}

\begin{document}

	\begin{center}
		{\LARGE {Orthogonal and symplectic
     Yangians  \\ and Yang-Baxter $R$-operators }}

 \vspace{0.3cm}

\large \sf A.P. Isaev$^{a}$\footnote{\sc e-mail:isaevap@theor.jinr.ru}, \
D. Karakhanyan$^{b}$\footnote{\sc e-mail: karakhan@yerphi.am},
R. Kirschner$^c$\footnote{\sc e-mail:Roland.Kirschner@itp.uni-leipzig.de} \\

\vspace{0.5cm}

\begin{itemize}
\item[$^a$]
{\it Bogoliubov Lab., Joint Institute of Nuclear Research, Dubna, Russia}
\item[$^b$]
{\it Yerevan Physics Institute,
 2 Alikhanyan br., 0036 Yerevan, Armenia }
\item[$^c$]
{\it Institut f\"ur Theoretische
Physik, Universit\"at Leipzig, \\
PF 100 920, D-04009 Leipzig, Germany}
\end{itemize}
\end{center}
\vspace{2 cm}
\begin{abstract}
\noindent
Yang-Baxter $R$ operators symmetric with respect to the orthogonal
and symplectic algebras are considered in an uniform way.
Explicit forms for the spinorial and metaplectic $R$ operators are obtained.
$L$ operators, obeying the $RLL$ relation with the orthogonal or symplectic
fundamental $R$ matrix, are considered in the interesting cases, where their
expansion in inverse powers of the spectral parameter is truncated.
Unlike the case of  special linear algebra symmetry
the truncation results in additional conditions on the Lie algebra
generators of which the $L$ operators is built and which can be fulfilled
in distinguished representations only.
Further, generalised  $L$ operators, obeying the modified $RLL$ relation
with the fundamental $R$ matrix replaced by the spinorial or metaplectic one, are
considered in the particular case of linear dependence on the spectral
parameter. It is shown how by fusion with respect to the spinorial or
metaplectic representation these first order spinorial $L$ operators
 reproduce the ordinary $L$ operators with second order truncation.
\end{abstract}


	\renewcommand{\refname}{References.}
	\renewcommand{\thefootnote}{\arabic{footnote}}
	\setcounter{footnote}{0}

\section{Introduction}

Let ${\cal G}$ be a Lie algebra of a Lie group $G$ and
$V_j$ be spaces of representations $\rho_j$ of
${\cal G}$ and $G$.
We consider the Yang-Baxter (YB) relations in the general form
\be \label{YB}
 		R_{12}(u)R_{13}(u+v)R_{23}(v)=
		R_{23}(v)R_{13}(u+v)R_{12}(u) \; \in \;
 {\rm End} (V_1 \otimes V_2 \otimes V_3 ) \; ,
\ee
 where the operator $R_{ij}$ acts nontrivially only in the
 spaces $V_i$ and $V_j$ and $u,v$ are spectral parameters.
It is  well known that (\ref{YB}) is the  basic relation in the treatment of integrable quantum systems
and is considered as an analog of the Jacobi identities in the
formulation of the related algebras \cite{FST,TTF,KuSk1,Fad,Drinfeld,Drin,ZZ} .

A solution $R_{ij}(u)$ of the YB relation (\ref{YB})
 is called  symmetric with respect to the group $G$ or the
algebra $\mathcal{G}$ if
the action of $R_{ij}(u)$ on $V_i\otimes V_j$ commutes with the
 action of the group $G$ (or its Lie algebra $\mathcal{G}$) in the
representation $\rho_i \otimes \rho_j$:
 $$
 [\rho_i(g) \otimes \rho_j(g) \, , \, R_{ij}(u)] = 0 \;\;\;\;
 (\forall g \, \in \, G) \;\; \Leftrightarrow \;\;
 [\rho_i(A) \otimes 1_j +  1_i \otimes \rho_j(A) \, ,
 \, R_{ij}(u)] = 0 \;\; (\forall A \, \in \, {\cal G}).
 $$
The present paper is concerned with
 the specific features of the YB relations
and the involved $R$ operators in the cases of symmetry with respect to
orthogonal ($so$) and symplectic ($sp$) algebra actions.
The less trivial representation theories in
those algebras compared to the special linear ($s\ell$) ones imply more involved
structures in the Yang-Baxter $R$ operators. A  distinguishing feature of
$so$ and $sp$ algebras compared to the $s\ell$ ones is the presence of an
invariant metric -- the second rank tensor $\varepsilon$,
determining the scalar product in the defining (fundamental)
representation. It is  symmetric, $\varepsilon^T = \varepsilon$,
 in the $so$ case and anti-symmetric, $\varepsilon^T = -\varepsilon$, in
the $sp$ case. In particular this results in the analogy
between the $so$ and $sp$ cases connected with the
interchange of symmetrisation with anti-symmetrisation
 and gives us the possibility to treat  both cases simultaneously.

The $so$ (or $sp$) symmetric  matrix $R_{ij}(u)$ obeying
 the Yang-Baxter (YB) relation (\ref{YB}),
 where $V_1=V_2=V_3$ are spaces of a defining (fundamental) representation,
is not a linear function in the spectral parameter $u$ as it is
for the $s\ell$ symmetric fundamental $R$-matrix. The explicit form of the
fundamental $so$ (and $sp$) symmetric  $R$-matrices
were  found first in \cite{ZZ,BKWK,Re}.

The generic YB relation (\ref{YB}) specifies to the $RLL$ relation
if
two of the three spaces $V_1 =V_2=V_f$ carry the fundamental representation $\rho_f$
while the third space $V_3 = V$ is the space of
any representation $\rho$ of ${\cal G}$.
 In this case the $R$-operator
  acting on the product $V_f \otimes V$ is called $L$-operator
 (or $L$ matrix).
 For the $so$ and $sp$ cases the $RLL$ version of the
YB relation involving the fundamental $R$ matrices \cite{ZZ,BKWK,Re}
 together with the $L$
operators of the form
 \be
 \lb{Lintr}
 L(u) = u {\bf 1} + \frac{1}{2}
 \rho_f(G^a_{\; b}) \, \rho(G^b_{\; a}) \; ,
 \ee
  does not hold
 for an arbitrary representation $\rho$ of the generators
$G^b_{\; a}$. The spinor representation $\rho_s$
of the orthogonal algebra with
 $\rho_s(G_{a b}) = M_{ab} = \frac{1}{4} [ \gamma_a, \gamma_b]$,
  where $\gamma_a$ are Dirac gamma-matrices,
  is a distinguished case, where the $RLL$ relation
is obeyed with $L$ of first order in the spectral parameter $u$
(see (\ref{Lintr})).
Also the spinorial $R$ matrix $R_{ss}(u)$, intertwining two spinor
representations $\rho_s$, is known \cite{SW}, \cite{ZamL}.
 This and other representations of the
orthogonal algebra distinguished in this sense
as well as the corresponding $R$ operators  (including spinorial $R$
operator) have been recently considered
 and analysed in detail in \cite{CDI,CDI13}.

 As we mentioned above we rely on the known similarity of the $so$ and $sp$ algebras
and treat the
related $R$ operators in a uniform way. In Section 2 we recall the fundamental $R$
matrices and present them for both cases uniformly.
Further  we identify the symplectic counterpart of the spinor representation.

 It is known that the $\mathcal{G}$-symmetric
 $RLL$ relations are defining relations for the
  infinite dimensional algebra called Yangian
$Y(\mathcal{G})$ of the type $\mathcal{G}$.
This concept was introduced by Drinfeld in \cite{Drin} and provides the
appropriate general viewpoint onto the known examples of simple forms
of operators $L(u)$ mentioned above. We use this concept
 to answer the question what are the general conditions for such simple
solutions to exist.
In general $L(u)$ obeying the $RLL$ relation expands in
inverse powers of the spectral parameter $u$ with infinitely many terms.
The truncation of the expansion of $L(u)$ at the first
non-trivial term (\ref{Lintr})
is known to be consistent for arbitrary representations
 $\rho$ of the $s\ell$ algebras
 (the evaluation representation of $Y(s\ell(n))$).
However, in the cases of $so$ and $sp$  symmetric $RLL$ relations the
truncation of the type (\ref{Lintr})
 results in additional conditions, which can be fulfilled only for
distinguished representations $\rho$.

 We shall investigate the additional conditions arising from the
truncation at the first and second order. The additional conditions appear
as characteristic equations in the matrix of generators
 $\rho_f(G^a_{\; b}) \, G^b_{\; a}$ which  enters  the definition
 of the $L$-operator. We stress here that
 a number of such examples has been considered in \cite{Re}.

In Section 5 the Yang-Baxter $R$ operator intertwining two spinorial
representaions and two metaplectic representations
is obtained in both orthogonal and symplectic cases  in the uniform  way.

It can be checked
that the  fundamental $R$ matrix (quadratic in $u$) can be reproduced by
fusion including projection from the product of the spinorial $L$ with its conjugate.
By  fusion of $R$ operators acting in the tensor product of the spinorial
and Jordan-Schwinger type representation spaces
we obtain the $L$ operator of second order in $u$ acting in $V_f \otimes V$ with $V$
carrying a representation of Jordan-Schwinger type, bosonic in the $so$ case
and fermionic in the $sp$ case.

\section{Fundamental  Yang-Baxter $R$-matrices}

Let $V_f$ be the space of the defining (fundamental) representation
of the Lie algebra ${\cal G}$ and its group $G$.
Let $V_f$ be the $n$-dimensional vector space with the basis vectors
$\vec{e}_a \in V_f$ $(a=1,\dots,n)$. Introduce the
 operator $R(u)$
which acts in the space $V_f \otimes V_f$
 \be
 \lb{Rope}
 R(u) \cdot (\vec{e}_{a_1} \otimes \vec{e}_{a_2}) =
 (\vec{e}_{b_1} \otimes \vec{e}_{b_2}) \, R^{b_1b_2}_{a_1a_2}(u) \; ,
 \ee
 and depends on the spectral parameter $u$. The
 matrix with elements $R^{b_1b_2}_{a_1a_2}(u)$
  is called fundamental Yang-Baxter (YB)
 $R$-matrix if it satisfies the Yang-Baxter equation of the form
	\be\label{ybe}
	\begin{array}{c}		R_{b_1b_2}^{a_1a_2}(u)R_{c_1b_3}^{b_1a_3}(u+v)
 R^{b_2b_3}_{c_2c_3}(v)
 =R_{b_2b_3}^{a_2a_3}(v)R_{b_1c_3}^{a_1b_3}(u+v)
 R^{b_1b_2}_{c_1c_2}(u) \; \Rightarrow \\ [0.2cm]
 R_{12}(u)R_{13}(u+v)R_{23}(v)=
 R_{23}(v)R_{13}(u+v)R_{12}(u) \; .
	\end{array}
	\ee
This equation is understood as a relation of
operators acting in $V_f \otimes V_f \otimes V_f$ and $R_{13}(u)$
or $R_{23}(u)$, etc.,
 denotes that the $R$-operator (\ref{Rope})
 acts nontrivially only in  first and third, or
 in second and third, etc.,
 factors of $V_f \otimes V_f \otimes V_f$.

The simplest solution to Yang-Baxter equation (\ref{ybe})
	is the Yang matrix
     \be
     \lb{yanGL}
  R^{a_1a_2}_{b_1b_2}(u) =
   u \delta^{a_1}_{b_1} \delta^{a_2}_{b_2} +
   \delta^{a_1}_{b_2} \delta^{a_2}_{b_1} = (u I + P)^{a_1a_2}_{b_1b_2} \; ,
   \ee
   Here $I$ and $P$ denote the unit and the permutation operators.
This YB $R$ matrix
is $g\ell(n)$ symmetric and acts in the tensor product of two fundamental
representations. The hierarchy of solutions
 of the Yang-Baxter equations, corresponding to higher
 representations can be obtained by the fusion method.
	
	In the orthogonal and symplectic cases the analoga
of the matrix (\ref{yanGL}) and the hierarchy of fusion
	solutions of  Yang-Baxter equations
 look more complicated and do not realize in the simplest
	way. To explain this we recall that operators $A$ acting in the
$n$-dimensional vector space $V_f$
 are elements of the algebra
 $so(n)$, or $sp(2m)$ $(2m=n)$, if the matrices
 $||A^a_{\;\; b}||_{a,b=1,\dots,n}$ of the operators $A$ in the basis
 $\vec{e}_a \in V_f$: $A \cdot \vec{e}_a = \vec{e}_b \, A^b_{\;\; a}$
  satisfy the conditions
 \be
 \lb{symb2}
 A^d_{\;\; a} \; \varepsilon_{db} + \varepsilon_{ad} \;  A^d_{\;\; b} = 0 \; ,
 \ee
 where $\varepsilon_{ab}$ is a non-degenerate
 invariant metric in $V_f$
  \be
  \lb{meSOSp}
   \varepsilon_{ab} = \epsilon \, \varepsilon_{ba}
 \; , \;\;\;  \varepsilon_{ab} \varepsilon^{bd} = \delta^d_a \; ,
 \ee
 which
 is symmetric $\epsilon = +1$ for $SO(n)$ case and skew-symmetric
 $\epsilon = -1$ for $Sp(n)$ case.
 We denote by $\varepsilon^{bd}$ (with  upper indices)
 the elements of the inverse matrix $\varepsilon^{-1}$.
	Namely the existence of the invariant tensor
	$\varepsilon_{ab}$ leads to the above mentioned complications in
  $SO(n)$ and $Sp(n)$ cases, e.g., it causes a third term in the
	corresponding expressions of the $R$-matrices and leads to the dependence
        on the
	spectral parameter of second power.
	
	The well known $R$-matrices
	\cite{ZZ,BKWK,SW,Re} for the $SO(n)$ and $Sp(n)$ $(n=2m)$ cases
   can be written in a unified form for arbitrary metrics $\varepsilon_{ab}$
   (\ref{meSOSp}) as follows (see, e.g., \cite{Isa})
	\be\label{rzW}
	R_{b_1b_2}^{a_1a_2}(u)=u(u + \frac{n}{2} -\varepsilon)
	I^{a_1a_2}_{b_1b_2}
	+(u + \frac{n}{2} -\varepsilon)P^{a_1a_2}_{b_1b_2}
	- \epsilon \, u \, K^{a_1a_2}_{b_1b_2} \; ,
	\ee
	where
 \be
 \lb{KP}
  I^{a_1a_2}_{b_1b_2} = \delta^{a_1}_{b_1} \delta^{a_2}_{b_2}   \; , \;\;\;\;
 P^{a_1a_2}_{b_1b_2} = \delta^{a_1}_{b_2} \delta^{a_2}_{b_1}   \; , \;\;\;\;
 K^{a_1a_2}_{b_1b_2} = \varepsilon^{a_1a_2} \, \varepsilon_{b_1b_2}  \; ,
 \ee
and the choices $\epsilon = +1$ and
$\epsilon = -1$ correspond to the $SO(n)$ and $Sp(n)$
 cases respectively. We note that the $R$-matrix (\ref{rzW}) is
 invariant under the adjoint action of any real form
 (related to the metric $\varepsilon^{ab}$) of the complex groups
 $SO(n,\mathbb{C})$ and $Sp(n,\mathbb{C})$.

Let the index range be $a_1,a_2,\dots = 1,\dots,n$  for the $SO(n)$ case and
	$a_1,a_2,\dots = -m,\dots,-1,1,\dots,m$ -- for the $Sp(n)$ $(n=2m)$ case.
  For the
 choice $\varepsilon^{a_1a_2} = \delta^{a_1a_2}$ in the $SO(n)$  case
 we have
	\be\label{rz}
	R_{b_1b_2}^{a_1a_2}(u)=u(u+\b)\delta^{a_1}_{b_1}\delta^{a_2}_
	{b_2}+(u+\b)\delta^{a_1}_{b_2}\delta^{a_2}_{b_1}-u\delta^
	{a_1a_2}\delta_{b_1b_2} \; , \;\;\;\; \b = (n/2-1) \; ,
	\ee
and for the choice
$\varepsilon^{a_1a_2} = \varepsilon_{a_2} \delta^{a_1, -a_2}$
(here $\varepsilon_{a} = sign(a)$ and
$\varepsilon^{ab} = - \varepsilon_{ab}$) in the $Sp(2m)$ case we have
	\be\label{rk}	R_{b_1b_2}^{a_1a_2}(u)=u(u+\b)\delta^{a_1}_{b_1}\delta^{a_2}_{b_2}
 +(u +\b)\delta^{a_1}_{b_2}\delta^{a_2}_{b_1} -
 u\e_{a_2}\e_{b_2}\delta^{a_1, -a_2}
 \delta^{b_1,-b_2}\; , \;\;\;\; \b = (m+1) \; .
	\ee
	
\section{Yangians of $so$ and $sp$ types}	

 Let ${\cal G}$ be the Lie algebra $so(n)$ or $sp(2m)$
 $(2m=n)$.
The Yangian $Y({\cal G})$ of
 ${\cal G}$-type is defined \cite{Drin} as an
associative algebra with the infinite number of
 generators $(L^{(k)})^a_b$ arranged as
 $(n \times n)$ matrices $||(L^{(k)})^a_b||_{a,b=1,...,n}$
  $(k=0,1,2,\dots)$ such that $(L^{(0)})^a_b = I \delta^a_b$,
 where  $I$ is an unit element in $Y({\cal G})$, and
  $(L^{(k)})^a_b$ for $k >0$ satisfy the quadratic defining relations which we
  shall describe  now.
   The generators
   $(L^{(k)})^a_b \in Y({\cal G})$ are considered as
   coefficients in the expansion of
\be
 \lb{defYan}
 L^a_b(u) = \sum_{k=0}^\infty  \frac{(L^{(k)})^a_b}{u^k}  \; , \;\;\;
 L^{(0)} = I \; ,
 \ee
 where $u$ is called spectral parameter.
The function $L(u)$ is called $L$-operator
and the defining relations of $Y({\cal G})$ are represented \cite{Drin}
as RLL-relations
 \be
 \label{rll}
 \begin{array}{c}
	R^{a_1a_2}_{b_1b_2}(u-v)L^{b_1}_{c_1}(u)L^{b_2}_{c_2}(v)=
	L^{a_2}_{b_2}(v)L^{a_1}_{b_1}(u)R^{b_1b_2}_{c_1c_2}(u-v) \;\;
 \Leftrightarrow \\ [0.2cm]
  R_{12}(u-v) \, L_1(u) \, L_{2}(v)=
	L_{2}(v) \, L_{1}(u) \, R_{12}(u-v)  \; .
  \end{array}
	\ee
Here $R^{a_1a_2}_{b_1b_2}(u-v)$  is the Yang-Baxter
 $R$-matrix (\ref{rzW}) and in the second line of (\ref{rll})
 we use the  standard matrix notations of \cite{FRT}.
   Recall that in this notations the
 matrices $R_{12}(u)$ (\ref{rzW}) and $P_{12}$, $K_{12}$ (\ref{KP})
 are operators in $V_f \otimes V_f$, where
 $V_f$ denotes a $n$-dimensional vector space, while $L_1$ and
 $L_2$  are matrices $||L^a_{\; b}||$ which
 act nontrivially only in the first and in the
 second factors of $V_f \otimes V_f$, respectively, and have algebra valued
 matrix elements.

 The defining relations (\ref{rll}) are homogeneous in $L$ and one can do in
 (\ref{defYan}), (\ref{rll}) the redefinition $L(u) \; \to \; f(u) \, L(u + b_0)$
 with any scalar function $f(u)= 1 + b_1/u + b_2/u^2 + \dots$, where
 $b_i$  are parameters.
 Now it is clear that
 the Yangian (\ref{rzW}), (\ref{rll}) possesses the set of automorphisms
 $$
 L(u) \; \to \; \frac{(u-a)^k}{u^k} \, L(u)  \; , \;\;\;\;
 \;\;\;\;\; (k =1,2,\dots) \; ,
 $$
 where $a$ is a constant (in general $a$ is a central element in $Y({\cal G})$).
 In particular for  $k=1$ we obtain that the generators
 $L^{(j)}$ are transforming as
 \be
 \lb{autY}
 \begin{array}{c}
 L^{(1)} \to L^{(1)} -a I_n \; , \;\;\;
  L^{(2)} \to L^{(2)} -a \, L^{(1)} \; , \;\;\; 
   L^{(3)} \to L^{(3)} - a L^{(2)} \; , \;\;\; \dots ,
   \end{array}
 \ee
 Taking $a = \frac{1}{n} {\rm Tr}(L^{(1)})$
 one can fix $L^{(1)}$ such that
 ${\rm Tr}(L^{(1)}) = 0$ (below we show that ${\rm Tr}(L^{(1)})$
 is central element in $Y({\cal G})$).

 We represent the fundamental $R$-matrix (\ref{rzW})
 in the concise form
 \be
 \lb{rzW1}
 R_{12}(u)=u(u + \beta) \, I
	+(u + \beta) \, P_{12}
	- \epsilon \, u \, K_{12} \; ,
 \ee
 where $\beta=(\frac{n}{2} -\epsilon)$. Further, after
 the shift of the spectral parameter $u \to u-v$,
  we write it as
 \be
 \lb{RRmat}
 \begin{array}{c}
 \frac{1}{u^2v^2}  R(u-v)=
  (\frac{1}{v}-\frac{1}{u})(\frac{1}{v} - \frac{1}{u}
 + \frac{\beta}{uv})- (\frac{1}{uv^2} - \frac{1}{u^2v}
 + \frac{\beta}{u^2v^2}) P -
 \epsilon (\frac{1}{uv^2} - \frac{1}{u^2v}) K  .
 \end{array}
 \ee
Then we substitute (\ref{RRmat}) and (\ref{defYan})
 into (\ref{rll}) and obtain, as a coefficient  at  $u^{-k}v^{-j}$,
  the explicit quadratic
 relations for the generators $(L^{(k)})^a_b$ of the Yangians
 $Y({\cal G})$
 \be
 \lb{defYan0}
 \begin{array}{c}
 [L_1^{(k)}, \; L_2^{(j-2)} ] -2 [L_1^{(k-1)}, \; L_2^{(j-1)} ] +
 [L_1^{(k-2)}, \; L_2^{(j)} ] \; + \\ [0.2cm]
 + \beta([L_1^{(k-1)}, \; L_2^{(j-2)} ] -
 [L_1^{(k-2)}, \; L_2^{(j-1)} ] ) \; + \\ [0.2cm]
 + \;
 P \Bigl(L_1^{(k-1)} \; L_2^{(j-2)}  -L_1^{(k-2)} \; L_2^{(j-1)}  +
 \beta L_1^{(k-2)} \; L_2^{(j-2)} \Bigr) \; - \\ [0.2cm]
 - \; \Bigl(L_2^{(j-2)} L_1^{(k-1)} -  L_2^{(j-1)}  L_1^{(k-2)} \;   +
 \beta  L_2^{(j-2)} \, L_1^{(k-2)} \Bigr) P \; + \\ [0.2cm]
  + \epsilon \Bigl(
  K\, (L_1^{(k-2)} \; L_2^{(j-1)}-L_1^{(k-1)} \; L_2^{(j-2)})  -
  (L_2^{(j-1)} L_1^{(k-2)}-L_2^{(j-2)} L_1^{(k-1)})\, K \Bigr)
  = 0  \; ,
 \end{array}
 \ee
 where the operators $K$, $P$
 are given in (\ref{KP}), $\epsilon = +1$ for ${\cal G} = so(n)$ and
 $\epsilon = -1$ for ${\cal G} = sp(2m)$.
For the special value $k=1$ we obtain from (\ref{defYan0})
 the set of relations
 \be
 \lb{defYan1}
 [ L^{(1)}_1 , \;  L^{(j-2)}_2] =
  - \left[ (P_{12} - \epsilon K_{12}) \, , \;  L^{(j-2)}_2 \right]  \; ,
  \;\;\;\;\; (\forall \; j) \; ,
 \ee
 which in particular lead to the statement that ${\rm Tr}(L^{(1)})$ is a central
 element in $Y({\cal G})$:
 $[ {\rm Tr}(L^{(1)}), \;  (L^{(j)})^a_{\;\; b}] =0$ $(\forall j)$.
 For $j=3$ we deduce from (\ref{defYan1})
 the defining relations for the Lie algebra generators
 $G^a_{\;\; b} \equiv -(L^{(1)})^a_{\;\; b}$:
  \be
   \lb{sospg}
 \begin{array}{c}
  [ G_1 , \;  G_2] =
  \left[ (P_{12} - \epsilon K_{12}) \, , \;  G_2 \right]  \; .
  \end{array}
 \ee
 The permutation of the indices $1 \leftrightarrow 2$ in this
 equation gives the consistency conditions and
 the same conditions are obtained from (\ref{defYan0}) directly.
 $$
 K_{12}\, ( G_1 + G_2 ) =
  ( G_1 + G_2 ) \,  K_{12}  \; ,
  $$
 Acting on this equation by $K_{12}$ from the left
   (or by $K_{12}$ from the right) we write it as
  \be
  \lb{symb0}
  K_{12}\, ( G_1 + G_2 ) = \frac{2}{n} \, {\rm Tr}(G) \, K_{12}
 =  ( G_1 + G_2 ) \, K_{12}  \; ,
  \ee
  where we have used
    \be
  \lb{idkn}
  K_{12}^2 = \epsilon \, n \, K_{12} \; , \;\;\;
  K_{12} \, N_1 \, K_{12} = K_{12} \, N_2 \, K_{12} =
  \epsilon \, {\rm  Tr}(N) \, K_{12} \; ,
  \ee
  Here $N$ is any $n \times n$ matrix.

  Then, according to the automorphism
  (\ref{autY}) we redefine the elements
   $G \to G - \frac{1}{n} \, {\rm Tr}(G)$
in  such a way  that for the new generators we have ${\rm Tr}(G) = 0$
 This leads
  to the (anti)symmetry conditions
   for the generators (cf. (\ref{symb2}))
 \be
   \lb{symb}
 \begin{array}{c}
  K_{12}\, ( G_1 + G_2 ) = 0 =
  ( G_1 + G_2 ) \,  K_{12}  \;\;\;\;\;\;
   \Rightarrow  \\ [0.2cm]
  G^{d}_{\;\;  a} \; \varepsilon_{db} +
  \varepsilon_{ad} \; G^{d}_{\;\;  b} = 0  \; .
  \end{array}
 \ee
 The equations
 (\ref{sospg}) and (\ref{symb}) for $\epsilon = +1$
 and $\epsilon = -1$  define the Lie
 algebra ${\cal G} =so(n)$ and ${\cal G}=sp(2m)$ $(2m=n)$,
  respectively.
  The defining
 relations (\ref{sospg}) and (anti)symmetry condition (\ref{symb})
for the generators $G_{ab} =
 \varepsilon_{ad} \; G^{d}_{\;\;  b}$
can be  written in the familiar form
 \be
 \lb{sospg1}
 \begin{array}{c}
 [ G_{ab}  , \; G_{cd} ] =
  \varepsilon_{cb} G_{ad} + \varepsilon_{db} G_{ca} +
 \varepsilon_{ca} G_{db} +
 \varepsilon_{da} G_{bc}  \;\; , \;\;\;\;\;\;
 G_{ab} =  - \epsilon \, G_{ba}  \;\; .
 \end{array}
 \ee
 This means (see \cite{Drin}) that
 an enveloping algebra
 ${\cal U}({\cal G})$ of the Lie algebra ${\cal G} = so(n),sp(2m)$
 is always a subalgebra in the Yangian $Y({\cal G})$.

 	\section{$L$ operators}

 Now we consider two  reductions  of the Yangian
 $Y({\cal G})$ (\ref{defYan0})
  which we call {\em linear and quadratic} evaluations, i.e.
  the two cases where $L(u)$ is represented
  by a linear or a quadratic polynomial in $u$.

 {\bf 1. Linear evaluation of $Y({\cal G})$.}

 We put equal to zero all generators
 $L^{(k)} \in Y({\cal G})$ with $k > 1$.
 In this case the $L$-operator (\ref{defYan}),
 after the  multiplication by $u$, is represented as
 \be \label{L}
  L^{a}_{\; b}(u) = u \delta^{a}_{b} - G^{a}_{\;\;  b} \; ,
	\ee
where for the Lie algebra generators we again use the notation
 $G^{a}_{\;\;  b}= -(L^{(1)})^{a}_{\;\;  b}$. It happens that for the choice
of the $L$-operator in the form (\ref{L}) the $RLL$
relations (\ref{rll}) in addition to the Lie algebra defining relations
 (\ref{sospg}) and (\ref{symb0}) lead to further
   constraints on the generators
 $G^{a}_{\;\;  b} \in {\cal G}$.

 \noindent
{\bf Proposition 1.} {\it For $so(n)$ and $sp(n)$
 type $R$-matrices (\ref{rzW1})
 the $L$-operator (\ref{L}) is a solution of (\ref{rll}) iff
 the elements $G^a_{\;\; b}$ satisfy
 (\ref{sospg}), (\ref{symb0})
 and in addition  obey the quadratic characteristic identity
 \be
 \lb{resosp-2a}
 \begin{array}{c}
 G^2  -  \left(\beta + \frac{2}{n} {\rm Tr} G \right) \, G -
 \frac{z}{n}  \, I_n = 0  \; , \\ [0.2cm]
 z \equiv {\sf C}^{(2)}  -
 \Bigl(\beta + \frac{2}{n} {\rm Tr} (G) \Bigr) \, {\rm Tr}(G) \; ,
 \end{array}
 \ee
 where $\beta=\frac{n}{2} - \epsilon$.
 The quadratic Casimir operator
 ${\sf C}^{(2)} = {\rm Tr}(G^2) = G^a_{\;\; b} \, G^b_{\;\; a}$
 is the central element in ${\cal U}({\cal G})$
  and $I_n$ is $n \times n$ unit matrix.} \\

{\bf Proof.}  Substitute the $R$-matrix (\ref{rzW1}) and the
  $L$-operator (\ref{L}) into (\ref{rll}).
  After a straightforward calculation we obtain
 that the $L$-operator (\ref{L}) is a solution of
 the equation (\ref{rll}) iff $G^a_{\;\; b} $
 satisfy the equations (\ref{sospg}), (\ref{symb0}) and
 \be
 \lb{resosp-0}
 K_{12} \; (G_1 \cdot  G_2 +
  \beta  \, G_2 ) = (G_2 \cdot  G_1 +
  \beta  \, G_2 ) \;  K_{12} \; ,
 \ee
   where $\beta = (\frac{n}{2} - \epsilon)$.  Note that the same
  condition (\ref{resosp-0}) can be obtained directly from the
  defining relations (\ref{defYan0}) of the Yangian  if we put there
  $L^{(k)} =0$, for $\forall k > 1$, and fix $j=2,k=3$
  (or $j=3,k=2$).
 Taking into account (\ref{symb0}) we write (\ref{resosp-0}) in the form
  \be
 \lb{resosp-1}
 \left[ K_{12} \, ,\; G_2^2 - \beta'
   \, G_2  \right] = 0 \; ,
 \ee
 where $\beta' = \Bigl( \beta + \frac{2}{n} {\rm Tr} (G)\Bigr)$.
 We act on the left hand side
  of (\ref{resosp-1}) by $K_{12}$ from the right
  and use formulas (\ref{idkn})
  which lead to the identities
  $$
  K_{12} \, G_2 \, K_{12} = \epsilon {\rm Tr} (G_2)
  \, K_{12}  \; , \;\;\;
  K_{12} \, (G_2)^2 \, K_{12} =
  \epsilon \, {\sf C}^{(2)} \, K_{12} \; ,
  $$
  where  ${\sf C}^{(2)} = {\rm Tr} (G_2)^2 =
  G^a_{\;\; b} \, G^b_{\;\; a}$.
As a result we have
  \be
 \lb{resosp-3}
  \begin{array}{c}
  K_{12} \,  (G_2^2 - \beta' G_2) \, K_{12} - \epsilon \, n \,
  (G_2^2 - \beta' \, G_2) \, K_{12} =   \\ [0.2cm]
  = \epsilon \Bigl( {\sf C}^{(2)} - \beta' \, {\rm Tr}(G)  -  n \, (G_2^2 -
  \beta' \, G_2) \Bigr) K_{12} = 0 \; ,
  \end{array}
  \ee
which is equivalent to (\ref{resosp-2a}). \hfill \qed

 \vspace{0.2cm}

 \noindent
 {\bf Remark 1.}  Since ${\rm Tr}(G)$ is a central element
  for the algebra (\ref{sospg}), one can shift
  the  spectral parameter $u \to u + \frac{1}{n} Tr(G)$
   in (\ref{rll}), (\ref{L}) and fix the generators $G^a_{\;\; b}$ such that
  $Tr(G)= G^a_{\;\; a}=0$. In this case, $G^a_{\;\; b}$ are generators of
  the Lie algebras ${\cal G}=so(n),sp(n)$ which satisfy (\ref{sospg}), (\ref{symb}).
  In this case the condition (\ref{resosp-2a})
  is simplified to
  \be
 \lb{resosp-2}
 \begin{array}{c}
 G^2  -  \left(\frac{n}{2} - \epsilon \right) \, G -
 \frac{1}{n} \, {\sf C}^{(2)} \, I_n = 0,
 \end{array}
 \ee
 that can
 be written also in the form
 $$
  G^a_{\;\; d} \, G^d_{\;\; b}   -
  \frac{1}{n}  \, \delta^a_b \,
  \Bigl( G^e_{\;\; d} \, G^d_{\;\; e} \Bigr)
  = \beta \, G^a_{\;\; b}   \; .
 $$
  The left hand side is
 the traceless quadratic combination
 of the matrices $G$ and the right hand side is
 proportional to the traceless matrix of generators
 $G^a_{\;\; b} \in {\cal G}=so(n),sp(2m)$ $(2m=n)$.

Writing $(G^2)^a_b$ as a sum of commutator and anti-commutator
we obtain
$$ (G^2- \beta G)^a_b = \half [G^a_c, G^c_b]_+. $$
This leads still to another form of the additional condition
(\ref{resosp-2}),
$$ [G^a_c, G^c_b]_+ = \frac{1}{n} {\sf C}^{(2)} \delta^a_b.
$$

 \vspace{0.2cm}

We stress that (\ref{resosp-2}) does not hold as an
 identity in the enveloping algebras ${\cal U}(so(n))$, or
  ${\cal U}(sp(n))$.
 This condition  can be fulfilled only when
 the generators $G^a_{\;\; d}$ are
 taken in some special representations of $so(n)$, or $sp(n)$.
 Thus, for the
 ansatz of the $L$-operator  (\ref{L})  we find that (\ref{rll})
	is valid only for a restricted class of
	representations of the Lie algebras $so(n)$ and $sp(2m)$.
 For the $so$ case this fact has already been noticed in \cite{Re,CDI,CDI13}
and for the $sp$ case it was discussed in \cite{Re}.
  We note also
 that the quadratic condition (\ref{resosp-2}) in the
 universal form (when the quadratic Casimir operator is not fixed)
 has been  discussed for the
 $so$ case in \cite{CDI13} and in the context of some
special representations of $so$ and $sp$ in \cite{Re}.
  Below we give examples of special  $so(n)$ and $sp(n)$
  representations which fulfill  the condition (\ref{resosp-2}).

\vspace{0.3cm}	

 \noindent
 {\bf Remark 2.} 	
	The definition of the of Yangian $Y({\cal G})$ of  ${\cal G}$-type
applies of course to case   of $g\ell_n$.
We take the $L$-operator and the $RLL$ relations in the same
form as in (\ref{defYan}), (\ref{rll}) and use in
(\ref{rll})  Yang's $R$-matrix (\ref{yanGL}). Thus,
 all relations for the Yangian $Y(g\ell_n)$ can be deduced from
(\ref{rzW1}),  (\ref{defYan0}) and (\ref{sospg})
if we put everywhere $K=0$ and $\beta=0$. For the
$g\ell_n$ case the analog
of the Proposition 1 states that the $L$-operator (\ref{L})
constructed from the set of elements
   $G^b_{\;\; c}$ obeying the $RLL$ relation (\ref{rll})
	with the fundamental Yang $R$-matrix (\ref{yanGL}) iff
elements $G^b_{\;\; c}$ satisfy the
defining relations for
 generators of $g\ell_n$ (cf. (\ref{sospg}) for $K=0$)
 $$
 [ G^{a_1}_{b_1} , \;  G^{a_2}_{b_2}] =
  \delta^{a_2}_{b_1} G^{a_1}_{b_2} - \delta^{a_1}_{b_2} G^{a_2}_{b_1} \; ,
 $$
 and there do not arise any
additional constraints (like (\ref{resosp-2})) on the generators
 $G^b_{\;\; c} \in g\ell_n$. It means that for $Y(g\ell_n)$ we
 have a homomorphic map of the Yangian
  $Y(g\ell_n)$ into the enveloping algebra
  ${\cal U}(g\ell_n)$ such that
  $$
  L^{(1)} \; \to \; - G \; , \;\;\;\;\;
  L^{(k)} \; \to \; 0 \;\;\;\;\;\; \forall k > 1 \; .
  $$
  This map is called {\em evaluation representation} of the Yangian
  $Y(g\ell_n)$.


 \vspace{0.2cm}

	Now we construct a representation
of the Lie algebra with the defining relations (\ref{sospg}), (\ref{symb})
which fullfil the condition
(\ref{resosp-2}). This distinguished representation
 is realized in terms of fermionic and bosonic oscillators.

 First we introduce
 an algebra ${\cal A}$ of fermionic or bosonic oscillators with
 generators $c^a$ $(a = 1,\dots,n)$ and the defining relations
 \be
 \lb{oscil}
 [ c^a, \, c^b ]_\epsilon \; \equiv \;
  c^a \, c^b + \epsilon \; c^b \, c^a = \varepsilon^{ab} \; .
 \ee
 Here for the $SO(n)$ case ($\epsilon=+1$)
 the elements $c^a$ are fermionic oscillators
and for the
 $Sp(n)$ $(n=2m)$ case $(\epsilon=-1)$
 the elements $c^a$ are bosonic oscillators.
The algebra ${\cal A}$ with the defining  relations
(\ref{oscil}) is covariant under the action
of $SO(n)$ or $Sp(n)$ group $c^a \to  U^a_{\; b} \, c^b$, where
$U \in SO(n)$ or $U \in  Sp(n)$. For the set of dual oscillators
$c_a = \varepsilon_{ab} \, c^b$ we obtain from (\ref{oscil}) the
following relations (cf. (\ref{oscil}))
 \be
 \lb{oscil-2}
 [ c_a, \, c_b ]_\epsilon \; \equiv \;
  c_a \, c_b + \epsilon \; c_b \, c_a = \varepsilon_{ba} = \epsilon \, \varepsilon_{ab}
  \;\;\; \Leftrightarrow \;\;\;
 c_a \, c^b + \epsilon \; c^b \, c_a = \delta^{b}_{a}  \; .
 \ee
 In particular we have
 $$
 c_a \, c^a = \epsilon \; c^a \, c_a  = \frac{1}{2} \varepsilon^{ab} (
 c_a \, c_b + \epsilon \; c_b \, c_a ) = \frac{n}{2}  \; .
 $$

 \vspace{0.2cm}
 \noindent
 {\bf Proposition 2.}
 {\it Let $c_a$ and $c^b$ be the generators of the oscillator
 algebra ${\cal A}$ with the defining relations (\ref{oscil}),
 (\ref{oscil-2}). The operators
 \be
 \lb{oscil-1}
 \begin{array}{c}
 \rho(G^a_{\; b}) =
 F^a_{\; b} = (c^a \, c_b - \frac{\epsilon }{2} \delta^a_b) =
  \frac{1}{2}(c^a \, c_b  - \epsilon \; c_b  \, c^a) =
    \varepsilon^{ad}  \, c_{[d}c_{b)}   =
    \varepsilon_{bd}  \, c^{[a}c^{d)}   \; , \\ [0.2cm]
     \rho(G^a_{\; a}) = F^a_{\; a} = {\rm Tr}(F) \; ,
    \end{array}
 \ee
 satisfy
 the relations (\ref{sospg}) and (\ref{symb}), i.e., the operators
 $F^a_{\; b} = \rho(G^a_{\; b})$
 define  representations $\rho$
 of $so(n)$ and $sp(n)$ generators.
 Moreover, the generators (\ref{oscil-1}) obey the quadratic
 characteristic identity
  \be
  \lb{char-2}
   F^{a}_{\;\; d} \; F^{d}_{\;\; b}
  - \beta \; F^{a}_{\;\; b} =
  \frac{1}{4} \, (n \epsilon -1) \, \delta^a_b  \; ,
 \ee
 which is nothing but the conditions (\ref{resosp-2}), where the
 quadratic Casimir operator is fixed as
 ${\sf C}^{(2)}=\frac{1}{4} n(n \epsilon -1) I$. Therefore
 the $L$-operator (\ref{L}) in the representtion (\ref{oscil-1})
 \be \label{L-1}
  L^{a}_{\; b}(u) = u \delta^{a}_{b} -  \varepsilon^{ad}  \, c_{[d}c_{b)} =
   u \delta^{a}_{b} -  \varepsilon_{bd}  \, c^{[a}c^{d)} \; ,
	\ee
 where $c_{[d}c_{b)} = \frac{1}{2}
 (c_d \, c_b  - \epsilon \; c_b  \, c_d)$ is the symmetrized
 product of $c_d$ and $c_b$, solves the $RLL$ relations (\ref{rll}).} \\

 {\bf Proof.} First, we check that the operators (\ref{oscil-1}) satisfy the
 symmetry condition (\ref{symb})
 $$
 \begin{array}{c}
 \Bigl( K_{12} (F_1 + F_2)
 \Bigr)^{a_1a_2}_{d_1b_2}  =
 K^{a_1a_2}_{d_1b_2} \, F^{d_1}_{\; b_1} +
 K^{a_1a_2}_{b_1d_2} \, F^{d_2}_{\; b_2} = \\
 = \varepsilon^{a_1a_2} \Bigl(
  (\epsilon c_{b_2} c_{b_1} + c_{b_1} c_{b_2})
  - \epsilon \,
  \varepsilon_{b_1b_2} \Bigr) =  0 \; .
 \end{array}
 $$
 Then after the substitution of (\ref{oscil-1})
 into the l.h.s. of (\ref{sospg}) we obtain
 $$
 \begin{array}{c}
 \left[ F^{a_1}_{\; b_1} \, , \;
  F^{a_2}_{\; b_2} \right] =
  \left[ c^{a_1} \, c_{\; b_1} , \;  c^{a_2} \, c_{\; b_2} \right] =
  \left[ c^{a_1} \, c_{\; b_1} , \;  c^{a_2} \right] c_{\; b_2} +
  c^{a_2} \, \left[ c^{a_1} \, c_{\; b_1} , \; c_{\; b_2} \right] =
  \\ [0.2cm]
  =
    \Bigl( c^{a_1} \left[  c_{b_1} , \;  c^{a_2} \right]_\epsilon -
 \epsilon \left[ c^{a_1} , \;  c^{a_2} \right]_\epsilon  c_{b_1}
  \Bigr) c_{b_2} +
  c^{a_2}\Bigl( c^{a_1} \left[  c_{b_1} , \; c_{b_2}
  \right]_\epsilon -
 \epsilon \left[ c^{a_1} , \;  c_{b_2} \right]_\epsilon  c_{b_1}
  \Bigr) = \\ [0.2cm]
  = \delta^{a_2}_{b_1} c^{a_1} c_{b_2} -
  \delta^{a_1}_{b_2} c^{a_2} c_{b_1} +
  \epsilon \bigl(c^{a_2} c^{a_1} \varepsilon_{b_1 b_2} -
  \varepsilon^{a_1 a_2} c_{b_1} c_{b_2}  \bigr) = \\ [0.2cm]
  = \delta^{a_2}_{b_1} \, F^{a_1}_{\; b_2} -
  \delta^{a_1}_{b_2} \, F^{a_2}_{\; b_1} +
  \epsilon \Bigl( F^{a_2}_{\; d_2} \;
  K^{a_1 d_2}_{b_1 b_2} -
  K^{a_1 a_2}_{b_1 d_2} \; F^{d_2}_{\; b_2}  \Bigr)
  \end{array}
 $$
 which is equivalent to (\ref{sospg}).  Finally we
 have
 $$
 \begin{array}{c}
 F^{a}_{\;\; d} \, F^{d}_{\;\; b}
  - \beta \; F^{a}_{\;\; b} =
  (c^a \, c_d - \frac{\epsilon }{2} \delta^a_d) (c^d \, c_b - \frac{\epsilon }{2} \delta^d_b) -
  \beta (c^a \, c_b - \frac{\epsilon }{2} \delta^a_b) = \\ [0.2cm]
  =(\frac{n}{2}  -\epsilon - \beta) c^a \, c_b
  + \frac{1}{4} \, \delta^a_b (1 + 2 \beta \epsilon ) =
  \frac{1}{4} \, \delta^a_b (n \epsilon -1)  \; ,
    \end{array}
 $$
 and the condition (\ref{resosp-2}) is fulfilled
for the value
$\rho({\sf C}^{(2)}) = \frac{1}{4} n(n \epsilon -1) I$ of the
quadratic Casimir operator.
 \hfill \qed

\vspace{0.3cm}	

 \noindent
 {\bf Remark 3.} 	
In the $so$ case the considered representation is formulated in terms of fermionic
oscillators. Their anti-commuation relation can be read also as the
defining relation of the Clifford algebra and one can identify the oscillator
generators with the Dirac gamma matrices.
$$ \gamma^a = \sqrt 2 c^a, \ \ [\gamma^a, \gamma^b]_+ = 2 \varepsilon^{ab}. $$
Therefore the representation is
called also spinorial. The bosonic oscillator representation is the
appropriate counterpart in the $sp$ case
and we shall call it metaplectic or also (symplectic) spinorial representation.
$$ \Gamma^a = \sqrt 2 c^a, \ \ [\Gamma^a, \Gamma^b]_- = 2 \varepsilon^{ab}. $$
The oscillator algebra and the spinor representation generators can be
regarded as appearing from the restriction of a Jordan-Schwinger type
representation of a general linear algebra \cite{KK15}.


  \vspace{0.2cm}

   {\bf 2. Quadratic evaluation of $Y({\cal G})$.}

   Now we put all generators
 $L^{(k)} \in Y({\cal G})$ with  $k > 2$ equal to zero.
 In this case the $L$-operator (\ref{defYan}),
 after a multiplication by $u^2$,
  can be written in the form
\be
 \lb{Lop2}
 L(u) = u(u+a) - u \, G + N \; ,
 \ee
where we introduce
 \be
 \lb{GN}
 L^{(1)}= a - G \; , \;\;\;\;  L^{(2)}= N \; ,
 \ee
 and $a$ is a constant.

We present  examples of distinguished representations allowing for a
quadratic evaluation of the Yangian of $so$ or $sp$ type and consider the
characteristic identities. The investigation of the additional condition in
the general case is subject of further study.


 \vspace{0.5cm}

Now we show that the fundamental representations $T=\rho_f$ of
$so$ and $sp$  is just an example of the quadratic evaluation of the
Yangian.

  \vspace{0.2cm}

 \noindent
 {\bf Proposition 3.}
 {\it The set of $n^2$ matrices
 $T(G^a_{\; b})$ (here indices $a$ and $b$
 enumerate matrices) with elements
 \be
 \lb{fund-1}
 T^c_{\; d}(G^a_{\; b})=
  - (P - \epsilon K)^{ac}_{\; b d} \equiv
 G^{ac}_{\; b d} \; ,
 \ee
 define the fundamental representation $T$ of
 generators $G^a_{\; b}$ for
 $SO(n)$ $\epsilon = +1$ and $Sp(n)$
 $\epsilon = -1$ cases.
 Generators (\ref{fund-1}) satisfy the cubic characteristic
 identity
   \be
  \lb{char-3}
 \begin{array}{c}
 G^{\, 3} + (1- n \epsilon) G^{\, 2}
 -  G = (1- n \epsilon) I \; ,
    \end{array}
 \ee
 The corresponding $L$-operator which
 solves the $RLL$ equation (\ref{rll})
 has the form (cf. (\ref{Lop2}))
 \be
 \lb{solFu}
 L(u) =  I + \frac{1}{u} (\beta \, I - G) +
 \frac{1}{2 u^2} \Bigl(G^2 - 2 \beta\, G - I \Bigr) \; ,
 \ee
 where $\beta = n/2-\epsilon$.}  \\
 {\bf Proof.} It is not hard to check that
 the generators (\ref{fund-1}) satisfy the conditions
  (\ref{symb}) and (\ref{sospg}) which in our
  notations (\ref{fund-1}) can be written in the form
 $$
 \begin{array}{c}
    K_{12} \, (G_{13} + G_{23}) = 0 =
   (G_{13} + G_{23}) \, K_{12}
    \; , \\ [0.2cm]
 [G_{13} , \; G_{23}] +
 [G_{12} , \; G_{23}] = 0 \; .
  \end{array}
 $$
Finally  for the matrix (\ref{fund-1}) we have the relations
 \be
 \lb{char-4}
 G^2 = I + (n \epsilon  -2) \; K \; , \;\;\;
G^3 = G + (n -2\epsilon ) \, (n \epsilon  -1) K \; ,
 \ee
 and the characteristic
 identity (\ref{char-3}) follows immediately from  (\ref{char-4}).

 We search the solution of (\ref{rll})
 as an $L$-operator acting in the space $V \otimes V$, where
 $V$ -- the space of fundamental representation of $so$ (or $sp$).
 In view of the Yang-Baxter equation (\ref{ybe}),
  it is clear that this $L$-operator is given
 by the $R$-matrix (\ref{rzW}) and can be represented in the form
 $$
 L(u)  = \frac{1}{u^2} \, R(u) = I + \frac{(\beta I - G)}{u} +
  \frac{\beta P}{u^2} =
  I + \frac{(\beta I - G)}{u}  +
 \frac{1}{2 u^2}\Bigl(G^2 - 2 \beta\, G - I \Bigr)  \;  ,
 $$
 where we have used the definition of $G$ (\ref{fund-1}) (written
 as $P = \epsilon K - G$) and the identities (\ref{char-4}).
\hfill \qed

 \vspace{0.2cm}

\noindent
{\bf Remark.} The quadratic Casimir operator in the representation (\ref{fund-1})
 is
 \be
 \lb{fund-2}
 \frac{1}{2} \; T^c_{\; d}({\rm Tr}(G^2)) =
 \frac{1}{2} \; (P - \epsilon K)^{ac}_{\; b r} (P - \epsilon K)^{br}_{\; a d}
 = (n - \epsilon) \delta^c_{\; d} \; .
 \ee
As a further example, consider the monodromy built  as the product of the
above spinor $L$ operators (\ref{L-1}) $ L_{12} (u) = L_1(u-\mu_1)
L_2(u-\mu_2)$ with multiplication in the fundamental representation space
$V_f$ and tensor product of two copies of the spinor space, i.e. it is
acting in $V_f\otimes V_{12}, V_{12} = V_s \otimes V_s$. The monodromy obeys
the $RLL$ relation and in this way we obtain an obvious example of the
second order evaluation of $Y({\cal G})$. Examples of higher order
evaluations are given by monodromies with more factors. In Sect. 6 we shall
consider the fusion procedure with projection from the tensor product of
spinor represetations to the fundamental one resulting in the fundamental
$R$ matrix.

 \vspace{0.5cm}

As the third example
 we consider the  representations of  $so$ and $sp$
of Jordan-Schwinger (JS) type where the
 generators are built from Heisenberg pairs in the form
 (cf. (\ref{oscil-1}))
 \be
 \lb{jshm01}
 M_{ab} = \epsilon(x_a \partial_b - \epsilon \, x_b \, \partial_a)=
   (\epsilon \, x_a \, \partial_b -x_b \, \partial_a) \; ,
 \ee
 where
 \be
 \lb{jshm02}
 \partial_a \, x_b - \epsilon \, x_b \, \partial_a =  \varepsilon_{ab}
 \; , \;\;\;
 \partial_a \, \partial_b = \epsilon \, \partial_b \, \partial_a
 \; , \;\;\;
 x_a \, x_b = \epsilon \, x_b \, x_a \; ,
 \ee
 and for $so$ and $sp$ we have $\epsilon = +1$ and $\epsilon = -1$
 respectively.  Contrary to the realization
 (\ref{oscil}), (\ref{oscil-2}), for the $so$ case
  $\{x_a,\partial_b\}$ are bosonic
 and for the $sp$ case  $\{x_a,\partial_b\}$ are fermionic. JS type
representations  have been considered for the $s\ell$ algebras e.g. in
\cite{KK09} and for the $so$ and $sp$ algebras in \cite {KK15}.
The defining relations and the (anti)symmetry condition are the same as in (\ref{sospg1})
 \be
 \lb{jshm03}
 [M_{ab} , \; M_{cd} ] = \varepsilon_{cb} M_{ad}
 + \varepsilon_{db} M_{ca} +
 \varepsilon_{ca} M_{db} + \varepsilon_{ad} M_{bc}\;\; , \;\;\;\;\;\;
  M_{ab}  = - \epsilon M_{ba} \; .
 \ee
 Let us introduce the operator
 $$
 H = (\varepsilon^{bc} x_b \, \partial_c ) =
 x_b \, \partial^b =
 \epsilon (\varepsilon^{bc} \partial_c x_b -n) \; ,
 $$
  which has the  properties
 $$
 H \, x_a = x_a \, (H +1) \; , \;\;\;\;
 H \, \partial_a = \partial_a \, (H - 1) \; .
 $$
Using (\ref{jshm01}) we find
 $$
  \begin{array}{c}
 (M^2)_{ad}  = M_{ab} M^b_{\;\; d} =
 \varepsilon^{bc} M_{ab} M_{cd} = \varepsilon^{bc} (\epsilon \, x_a \, \partial_b-x_b \, \partial_a )
 (\epsilon \, x_c \, \partial_d -x_d \, \partial_c) = \\ [0.2cm]
 =(\epsilon n -2) \, x_a \, \partial_d + \varepsilon_{ad}
 \, H + (H-1) (x_a \, \partial_d  + \epsilon x_d \, \partial_a )
 - x_a x_d \partial_b \partial^b - x_b x^b \partial_a \partial_d =  \\ [0.2cm]
 =(\epsilon n + 2 H -4) \, x_a \, \partial_d + \varepsilon_{ad}
 \, H - \epsilon (H-1) M_{ad}
 - x_a x_d \partial_b \partial^b - x_b x^b \partial_a \partial_d \; ,  \end{array}
 $$
  and  for
  $so$ and $sp$ Lie algebras in the considered representations (\ref{jshm01}))
  we have the following explicit form of the quadratic Casimir operator
 \be
 \lb{kaz2}
 {\rm Tr} (M^2) =   M^d_{\;\; b} M^b_{\;\; d} =
 \varepsilon^{da} \, (M^2)_{ad} = 2(n - 2\epsilon) \, H
 + 2 \epsilon H^2 - 2 \epsilon x^2 \partial^2 \; ,
 \ee
 where $x^2 = x_b x^b$, $\partial^2 = \partial_d \partial^d$.

In  the $so$ case we have here the
  finite dimensional representations of
  integer spins $m$ and they are spanned  by the homogeneous harmonic
polynomials $P_m(x)$
 $$
 \partial^2 \, P_m(x) = 0 \; , \;\;\; H \, P_m(x) = m \; P_m(x) \; .
 $$
 Then by using (\ref{kaz2}) we obtain
 \be
 \lb{specM2}
 \frac{1}{2} {\rm Tr}(M^2) \; P_m(x) = \Bigl( (n - 2\epsilon) \, m
 + \epsilon \, m^2 \Bigr) \; P_m(x) \; .
 \ee
 In  spin m=1 case we obtain the eigenvalue of the Casimir operator
 as $(n-\epsilon)$ which  coincides with the value for the
fundamental representation (\ref{fund-2}) as expected.

 \vspace{0.2cm}

\noindent
 {\bf Proposition 4.} {\it In the JS type representation  (\ref{jshm01})
 the characteristic identity for the generators of
 $so$ and $sp$ algebras is }
 \be
 \lb{char-3g}
 \begin{array}{c}
 M^3_{ad} = (n- \epsilon) M^2_{ad}   + (2 -\epsilon n) M_{ad}
 - \frac{1}{2} {\rm Tr}(M^2) (\varepsilon_{ad} - \epsilon M_{ad})
 \; , \\ [0.2cm]
 M^3 + (\epsilon-n) M^2   + (\epsilon n-2) M
 + \frac{1}{2} {\rm Tr}(M^2) (I - \epsilon M) = 0
 \; .
  \end{array}
 \ee
 {\bf Proof.} Multiply the matrix $M^2$ as written above by $M$. After straightforward
 calculations we get (\ref{char-3g}). \hfill \qed

 \vspace{0.2cm}

 Now we compare this formula (\ref{char-3g})
with the characteristic identity (\ref{char-3}) found for
 the case of the fundamental (defining) representation (\ref{fund-1}).
First we note that (\ref{char-3}) is transformed
 to (\ref{char-3g}) if we redefine $G \to \epsilon M$.
Then (\ref{char-3g}) gives (\ref{char-3}) for the choice of the value of the
Casimir operator
 $$
 \frac{1}{2} {\rm Tr}(M^2) = (n - \epsilon) \; ,
$$
 which is compatible with the spectrum (\ref{specM2}) for $m=1$
 and with (\ref{fund-2}).
	
The $L$-operator can be wrtten in the form

\be \label{LJS}
L(u) = (u-\lambda) (u-\mu) I + (u-\sigma) M + M^2
\ee

In Section 6 we shall show that this form is obtained by fusion from YB
operators acting in the product of the spinorial and the JS representation.
It can be checked that the $RLL$ relation with the fundamental $R$ matrix
(\ref{rll}) is fulfilled for particular values of the parameters.


\section{Spinorial and metaplectic Yang-Baxter operator $\check{\mathfrak{R}}$}
\setcounter{equation}{0}

 Let ${\cal A}$ be the algebra of
 (fermionic or bosonic) oscillators with the defining relations (\ref{oscil})
and denote by ${\cal G}$ the Lie algebra $so(n)$ or $sp(n)$.
  Consider the  $L$-operator in the general form
 \be
 \lb{L-2}
 L(u) = u \; I + \frac{1}{2} \, \rho(G^{a}_{\; b}) \otimes G^{b}_{\; a}  =
  u \; I - \frac{1}{2} \, c^{[a} \, c^{b)} \otimes G_{ab}
  \;\; \in \;\; {\cal A} \otimes {\cal U}_{\cal G} \; ,
 \ee
 where $u$ is the spectral parameter,
 $G^{b}_{\; a}$ are the  generators of the Lie algebra ${\cal G}$,
 ${\cal U}_{\cal G}$ denotes the enveloping algebra of ${\cal G}$.
 $\rho(G^{a}_{\; b}) \in {\cal A}$ denote the image of the generators
 $G^{b}_{\; a} \in {\cal G}$ in the
 oscillator representation (\ref{oscil-1}). Note that if we
 evaluate the second factor in (\ref{L-2}) in the fundamental representation
 (\ref{fund-1}) then the $L$-operator (\ref{L-2}) takes the form (\ref{L-1}):
 $$
 L^a_{\; b}(u) = u \; \delta^a_b +
 \frac{1}{2} \, \rho(G^{c}_{\; d}) \; T^a_{\; b}(G^{d}_{\; c}) =
 u \; \delta^a_b +  \frac{1}{2} \, \rho(G^{c}_{\; d})
 (\epsilon K^{da}_{\; cb} - P^{da}_{\; cb}) =
 u \; \delta^a_b - \rho(G^{a}_{\; b}) \; ,
 $$
 where we have used the symmetry properties (\ref{sospg1})
 of generators $G^{c}_{\; d}$.

 Now we consider the new version of the $RLL$ relation (cf. (\ref{rll}))
 \be
 \lb{rll-osc}
 \begin{array}{c}
\check{\mathfrak{R}}_{12}(u) \; L_1(u+v) \; L_2(v)=
 L_1(v) \; L_2(u+v) \; \check{\mathfrak{R}}_{12}(u) \; , \;\;
\end{array}
 \ee
different from  (\ref{rll})) because  $\check{\mathfrak{R}}(u) $ is not the
fundamental $R$ matrix, but rather an element of the algebra ${\cal A}
\otimes {\cal A} $,
 $\check{\mathfrak{R}}(u)=
 {\sf P} \cdot \mathfrak{R}(u) \in {\cal A} \otimes {\cal A} $.
 Operator ${\sf P}$ permutes the factors in the tensor
product ${\cal A} \otimes {\cal A}$ and the operators
 $\check{\mathfrak{R}}_{12}(u)$, $L_1$, $L_2$  are elements of
 $ {\cal A} \otimes {\cal A} \otimes {\cal U}_{\cal G}$:
 $\; \check{\mathfrak{R}}_{12}(u) = \check{\mathfrak{R}}(u) \otimes
 I_{{\cal U}_{\cal G}}$,
 \be \label{L-2sp}
 \begin{array}{c}
 L_1(v) = v \, I -
 \frac{1}{2} \, c^{[a} \, c^{b)} \otimes I_{\cal A} \otimes G_{ab}
 \equiv v \, I -
 \frac{1}{2} \, c_1^{[a} \, c_1^{b)} \, G_{ab} \; , \\ [0.2cm]
  L_2(v) = v \, I -
 \frac{1}{2} \, I_{\cal A} \otimes c^{[a} \, c^{b)} \otimes  G_{ab}
 \equiv v \, I -
 \frac{1}{2} \, c_2^{[a} \, c_2^{b)} \, G_{ab} \; .
 \end{array}
 \ee
 Here $I_{\cal A}$ is the  unit element in ${\cal A}$ and
 $I$  is the unit element in  ${\cal A } \otimes {\cal A } \otimes  {\cal U}_{\cal G}$
 We introduce the short-hand notations $c^a_1$ for oscillators $c^a$
 in the first factor of ${\cal A } \otimes {\cal A } \otimes  {\cal U}_{\cal G}$
 and $c^a_2$ in the second factor of
 ${\cal A } \otimes {\cal A } \otimes  {\cal U}_{\cal G}$.

Since the $L$-operator is fixed in (\ref{L-2}),
we interpret (\ref{rll-osc}) as the defining equation
 for the operator  $\check{\mathfrak{R}}_{12}(u)$.

 \vspace{0.2cm}

 Introduce the basis in the algebra ${\cal A}$ of fermionic or bosonic oscillators
 which is formed by the unit element $I_{\cal A}$ and the (anti-)symmetrized products
 \be
 \lb{basis}
 c^{[a_1}\cdots c^ {a_k)}  = \sum_{\sigma \in S_k} (-\epsilon)^{p(\sigma)} \;
  c^{a_{\sigma(1)}} \cdots c^{a_{\sigma(k)}}  \;\;\;\;\; (k=1,2,\dots) \; ,
 \ee
 where $S_k$ is a symmetric group,
 the sum is performed over all permutations $\sigma \in S_k$
 of $k$ indices $(1,2,\dots,k)$ and $p(\sigma)$ is the parity of $\sigma$.
  Under the transposition of any two indices $a_i$ and $a_j$, the basis elements
 (\ref{basis}) have the (anti)symmetric property
 $$
 c^{[a_1} \cdots c^{a_i} \cdots c^{a_j} \cdots c^{a_k)} = - \epsilon \;
 c^{[a_1} \cdots c^{a_j} \cdots c^{a_i} \cdots c^{a_k)} \; .
 $$

Preparing the proof of the following theorem
where we encounter the products of elements
of the (anti-)symmetrised basis we study
the expansion of such products
into the elements of this (anti-)symmetrised basis. This is
 conveniently done by using
generating functions (see, e.g., \cite{CDI13}). We introduce
the  auxiliary (anti-)commuting
 variables $\kappa^a,\kappa^{\prime a}, \dots$ such that
 $$
 \begin{array}{c}
  \kappa_a = \varepsilon_{ab} \, \kappa^b \; , \;\;\;
 \kappa^a \, \kappa^b = - \epsilon \kappa^b \, \kappa^a \; , \;\;\;
  \kappa^a \, \kappa^{\prime b} = - \epsilon \kappa^{\prime b} \, \kappa^a \; ,
  \\ [0.2cm]
  \kappa^a \, c^b = - \epsilon c^b \, \kappa^a \; , \;\;\;
    \kappa^{\prime a} \, c^b = - \epsilon \, c^b \, \kappa^{\prime a} \; , \;\;\;
  \end{array}
 $$
 and define the operators $\dd^b = \partial/\partial \kappa_b$,
 $\dd^{\, \prime b} = \partial/\partial \kappa'_b$, with
 relations
 \be
 \lb{dkappa}
 \begin{array}{c}
 [\dd^b , \; \kappa_a]_\epsilon  =
 [\dd^{\prime b} , \; \kappa'_a]_\epsilon  = \delta^{b}_a
  \; , \;\;\;
  [\dd_b , \; \kappa^a]_\epsilon  =
  [\dd'_b , \; \kappa^{\prime a}]_\epsilon  = \epsilon \; \delta^{a}_{b}
  \; , \\ [0.2cm]
  [\dd^b , \; \kappa^a]_\epsilon  =
  [\dd^{\prime b} , \; \kappa^{\prime a}]_\epsilon  = \varepsilon^{a b}
  \; ,  \;\;\;
  [\dd^{\prime b}, \; \dd^a]_\epsilon =
  [\dd^{\prime b}, \; \kappa_a ]_\epsilon =
  [\dd^{b}, \; \kappa'_a ]_\epsilon = 0 \; ,
  \end{array}
 \ee
 where $[A, \; B]_\epsilon = A \, B + \epsilon \, B \, A$
 and according to our agreement we have $(\dd_i)_b = \varepsilon_{ba} \dd_i^a =
 \epsilon \partial/\partial \kappa_i^b$.
 The scalar products of variables $\kappa^b$
 with themselves and with generators of ${\cal A}$ are
 $$
  \begin{array}{c}
 (\kappa \cdot \kappa') \equiv \kappa_a \kappa^{\prime a} =
\varepsilon_{ab} \, \kappa^b \kappa^{\prime a} =
 \epsilon \, \kappa^a \kappa'_a  = - \kappa'_a \kappa^a =
 -  (\kappa' \cdot \kappa) \; , \\ [0.2cm]
 (\kappa\cdot c)= \kappa_a c^a
 = - c_a \kappa^a
  = -  (c \cdot \kappa) \; , \;\;\; (\kappa \cdot \kappa) = 0 \; .
\end{array}
$$
 Thus, the variables $\kappa^a$, $\kappa^{\prime a}$ have the same grading as $c^a$'s,
i.e., they are
anti-commuting variables in the $so$ case ($\epsilon=+1$) and commuting ones in
the $sp$ case ($\epsilon=-1$). Then derivatives of the expression $(\kappa\cdot c)^k$
with respect to $\dd^{a_k}\ldots\dd^{a_1}$
will give the elements  of the symmetrized basis of ${\cal A}$:
$$
\dd^{a_1}\ldots\dd^{a_k} \,
(\kappa\cdot c)^k = k! \; c^{[a_1}\ldots c^{a_k)}  \; ,
$$
and it also can be written in the form
$$
c^{[a_1}\ldots c^{a_k)}=\dd^{a_1}\ldots\dd^{a_k}\ e^
{(\kappa\cdot c)} |_{\kappa=0} \; .
$$
  Then we consider the product of two
 basis elements of algebra ${\cal A}$ (see (\ref{Rdef}))
$$
 \left.
c^{[a_1}\ldots c^{a_k)} \cdot c^{[a}c^{b)}=\dd^{a_1}\ldots\dd^{a_k}
 \, e^{(\kappa\cdot c)}\
 \dd^{\, \prime a}\dd^{\, \prime b} \, e^{(\kappa' \cdot c)}
 \right|_{\kappa=\kappa'=0} \; =
$$
$$
\left.
=\dd^{a_1 \ldots a_k}
 \,  \dd^{\prime a b} \, e^{(\kappa \cdot c)} \ e^{(\kappa' \cdot c)} \right|_{\kappa=\kappa'=0}  =
 \left.
\dd^{a_1 \ldots a_k} \,  \dd^{\, \prime a b}
 \ e^{((\kappa+\kappa')\,\cdot\, c)}
 e^{(\frac\epsilon2 \kappa' \cdot\kappa)}
 \right|_{\kappa=\kappa'=0} \; .
$$
Here we denoted
$\dd^{a_1 \ldots a_k} \equiv \dd^{a_1}\ldots\dd_i^{a_k}$ $(i=1,2)$,  and
used the Baker-Hausdorff formula
$$
\begin{array}{c}
e^{(\kappa\cdot c)} \ e^{(\kappa' \cdot c)} =
e^{(\kappa\cdot c)+(\kappa' \cdot c) +
 \frac12 [(\kappa\cdot c),  \, (\kappa'\cdot c)]} \; , \\ [0.2cm]
 [(\kappa\cdot c),(\kappa'\cdot c)]=
 \kappa'_{b} \kappa_{a}[c^a,c^b]_{\epsilon}=
 \kappa'_{b} \kappa_{a} \, \varepsilon^{ab} = \epsilon
\; (\kappa' \cdot\kappa) \; .
\end{array}
$$
Then we change the variables $\{ \kappa, \, \kappa' \}$ to
$\{ \bar{\kappa} = \kappa +
\kappa', \, \kappa' \}$ and it leads to $(\kappa'\cdot\kappa) = (\kappa'\cdot\bar{\kappa})$
while the $\kappa$-derivatives are transformed as following
$$
\dd \to \bar{\dd}, \ \dd' \to \bar{\dd} + \dd' \; .
$$
Finally this change of variables results in
(for simplicity we remove bar for variables $\bar{\kappa}$, etc.)
$$
\left.
c^{[a_1}\ldots c^{a_k)}c^{[a}c^{b)}=\dd^{a_1\ldots a_k}(\dd^a+
\dd^{\prime a})(\dd^b+\dd^{\prime b})\
 e^{(\kappa\cdot c)}e^{ \frac\epsilon2(\kappa' \cdot
	\kappa)} \right|_{\kappa=\kappa'=0}=
$$
 $$
 \left.
=\dd^{a_1 \ldots a_k}[(\dd^a \dd^b e^{(\kappa\cdot c)}
-\epsilon \, \dd^b e^{(\kappa\cdot c)} \dd^{\prime a}
 + \dd^a e^{(\kappa\cdot c)} \dd^{\prime b}  + e^{(\kappa\cdot c)}
 \dd^{\prime a} \dd^{\prime b}  ]\
 e^{ \frac\epsilon2 (\kappa' \cdot
	\kappa)} \right|_{\kappa=\kappa'=0}=
$$
  \be\label{aab}
 \left.
=\dd^{a_1 \ldots a_k} \left[ \dd^a\dd^b
+ \frac12( \epsilon \kappa^{a}\dd^b -
\kappa^{b}\dd^a ) +\frac14 \kappa^{a}\kappa^{b}\right]
\ e^{(\kappa\cdot c)} \right|_{\kappa=0} \; .
 \ee
In the same way we obtain
$$
 \left.
c^{[a}c^{b)}c^{[a_1}\ldots c^{a_k)}=
 \dd^{\prime a}\dd^{\prime b} \, e^{(\kappa' \cdot c)} \,
  \dd^{a_1}\ldots\dd^{a_k}
 \, e^{(\kappa\cdot c)} \right|_{\kappa=\kappa'=0} \; =
$$
$$
\left.
=\dd^{a_1 \ldots a_k}
 \,  \dd^{\prime a b} \, e^{(\kappa' \cdot c)} \ e^{(\kappa\cdot c)} \right|_{\kappa=\kappa'=0}  =
 \left.
\dd^{a_1 \ldots a_k} \,  \dd^{\prime a b}  \ e^{((\kappa+\kappa'
)\,\cdot\, c)}e^{-\frac\epsilon2(\kappa'\cdot\kappa)} \right|_{\kappa=\kappa' =0} =
$$
$$
 \left.
=\dd^{a_1 \ldots a_k}[(\dd^a \dd^b e^{(\kappa\cdot c)}
-\epsilon \, \dd^b e^{(\kappa\cdot c)} \dd^{\prime a}
 + \dd^a e^{(\kappa\cdot c)} \dd^{\prime b}  + e^{(\kappa\cdot c)} \dd^{\prime a}  \dd^{\prime b}   ]\
 e^{ - \frac\epsilon2 (\kappa'\cdot \kappa)} \right|_{\kappa=\kappa'=0}=
$$
 \be\label{aba}
 \left.
=\dd^{a_1 \ldots a_k} \left[ \dd^a\dd^b
-  \frac12( \epsilon \kappa^{a}\dd^b -
\kappa^{b}\dd^a ) +\frac14 \kappa^{a}\kappa^{b}\right]
\ e^{(\kappa\cdot c)} \right|_{\kappa=0} \; .
 \ee
 The difference between expressions (\ref{aab}) and (\ref{aba})
 appears only in the sign of terms which are linear in $\kappa^d$. So we denote
 \be
 \lb{plumi}
 [\pm]^{ab}_i = \left[ \dd^a_i \dd^b_i
\pm \frac12( \epsilon \kappa^{a}_i\dd^b_i -
\kappa^{b}_i\dd^a_i ) +\frac14 \kappa^{a}_i\kappa^{b}_i\right] \; ,
 \ee
where the index $i$ refers the first or second factor in the tensor product
${\cal A} \otimes {\cal A} $.

\noindent
 {\bf Proposition 6.} {\em The $so$ or $sp$ invariant $R$-operator
 $\check{\mathfrak{R}}_{12}(u)$ which satisfies the $RLL$-relations
 (\ref{rll-osc}) with the $L$-operator given in (\ref{L-2}) has the form
 \be \label{Rrkosp-1}
\check{\mathfrak{R}}=\sum_k\frac{r_k(u)}{k !} \sum_{\vec{a},\vec{b}} \;
\e_{a_1b_1}\cdots \e_{a_kb_k} \; c^{[a_1}
\ldots c^{a_k)}\otimes c^{[b_1}\ldots c^{b_k)} \; ,
\ee
where $\e_{a b}$  is the $so$ or $sp$ invariant metric and the
coefficient functions $r_k(u)$ are
 written separately for even and odd $k$ as
 \be
 \lb{Rdef8}
 \begin{array}{c}
 \displaystyle{
  r_{2m}(u) = \frac{2^{2m} \, \Gamma(m  + \epsilon \frac{u}{2})}
  {\Gamma(m +1  - \epsilon \frac{(u+n)}{2})} A_0(u) \; , }
  \\ [0.4cm]
  \displaystyle{
  r_{2m+1}(u) =  \frac{2^{2m} \, \Gamma(m  + \frac{1}{2}+ \epsilon \frac{u}{2})}
  {\Gamma(m +\frac{1}{2} - \epsilon \frac{(u+n)}{2})} A_1(u) \; , }
  \end{array}
 \ee
 Here $A_0(u)$, $A_1(u)$ are  arbitrary functions and
 $\epsilon = +1$ for the  $so$ case and $\epsilon = -1$ for the $sp$ case.} \\

\vspace{.5cm}

 \noindent
 {\bf Proof.}

{\it 1st part: Extracting the defining conditions from the $RLL$ relation}

After the substitution of
 (\ref{L-2}) into (\ref{rll-osc}) the quadratic parts in spectral parameters
 (proportional to $v^2$ and $u v$) in both sides
 of (\ref{rll-osc}) are canceled. The linear
in $v$ term gives the symmetry condition
 \be
 \lb{sym}
[ ( \rho(G^a_{\; b}) \otimes I_{\cal A} +
I_{\cal A} \otimes \rho(G^a_{\; b}) ) , \; \check{\mathfrak{R}}(u)]   =  0
 \;\;\; \Leftrightarrow  \;\;\;
 [ ( c_1^{[a} c_1^{b)} +  c_2^{[a} c_2^{b)}  )  ,
  \; \check{\mathfrak{R}}(u)]   =  0 \; ,
 \ee
 which indicate that $\check{\mathfrak{R}}(u)$ is invariant under
 the adjoint action of $so$ and $sp$ algebras.
The terms in (\ref{rll-osc})
 which contains no $v$ leads to the
 condition:
\be \label{defR1}
 \begin{array}{c}
 u \Bigl(\check{\mathfrak{R}}_{12}(u) \; c_2^{[a}\, c_2^{b)} -
 c_1^{[a}\, c_1^{b)} \; \check{\mathfrak{R}}_{12}(u) \Bigr) G_{ab} - \\ [0.2cm]
  - \displaystyle {\frac{1}{2} } \,
  \Bigl(\check{\mathfrak{R}}_{12}(u) \; c_1^{[a}\, c_1^{e)} \; c_2^{[f} \, c_2^{b)}
   -  c_1^{[a} \, c_1^{e)} \; c_2^{[f} \, c_2^{b)}
   \; \check{\mathfrak{R}}_{12}(u) \Bigr) G_{ae} G_{fb}  = 0  \; .
\end{array}
\ee
Now we substitute
 $G_{ae} G_{fb} = \frac{1}{2} \bigl( [G_{ae}, \; G_{fb}]_{-} +
 [G_{ae}, \; G_{fb}]_{+}  \bigr)$
 and use commutation relations (\ref{sospg1}). Then the condition (\ref{defR1})
 is written in the form
 \be \label{defR2}
 \begin{array}{c}
\Bigl( u \bigl[ \check{\mathfrak{R}}_{12}(u) \; c_2^{[a b)} -
 c_1^{[ab)} \; \check{\mathfrak{R}}_{12}(u) \bigr]
  - \varepsilon_{fe} \, X^{[ae)[fb)} \Bigr) \; G_{ab}
   = \frac{1}{4} \, X^{[ae)[fb)}\, [G_{ae}, \; G_{fb}]_+    \; ,
\end{array}
\ee
where we denote $c^{[a b)} = c^{[a} \, c^{b)}$ and
$$
X^{[ae)[fb)} =
  \bigl( \check{\mathfrak{R}}_{12}(u) \, c_1^{[ae)} \, c_2^{[fb)}
   -  c_1^{[ae)} \, c_2^{[fb)}
   \, \check{\mathfrak{R}}_{12}(u) \bigr)
$$
Following \cite{CDI13} we search the solution of (\ref{defR2}) as the
solution of two equations
 \be \label{defR}
 \begin{array}{c}
\Bigl( u \bigl[ \check{\mathfrak{R}}_{12}(u) \; c_2^{[a b)} -
 c_1^{[ab)} \; \check{\mathfrak{R}}_{12}(u) \bigr]
  - \varepsilon_{fe} \, X^{[ae)[fb)} \Bigr) \; G_{ab} = 0   \; ,
\end{array}
\ee
 \be \label{defR3}
 X^{[ae)[fb)}\, [G_{ae}, \; G_{fb}]_+  = 0  \; .
\ee
Below it will be shown that equation (\ref{defR3}) leads to the condition
\be \label{defR5}
 [G_{[ae} G_{f)b}]_+ = 0  \; ,
\ee
where $[aef)$ denotes the (anti)symmetrization over three indices. This condition
is not valid for the enveloping
algebra ${\cal U}_{\cal G}$. Therefore we need to restrict our
 consideration to appropriate
representations of ${\cal U}_{\cal G}$ for which (\ref{defR5}) is fulfilled.
 Note that the operators $G_{ab}$ with symmetric property
 $G_{ab} = - \epsilon G_{ba}$ (cf. (\ref{sospg1})) satisfy the identity
 $[G_{[ae} G_{f)b}]_+ = [G_{a[e} G_{fb)}]_+$ and thus
 (\ref{defR5}) is valid for (anti)symmetrization of any
 three of four indices $(a,e,f,b)$.
 For the $so$ case the condition (\ref{defR5})
 was discussed  in \cite{CDI13}.

The Yangian-type condition (\ref{defR}) fixes the operator $\check{\mathfrak{R}}$
and we start to solve it now. We apply the method developed in \cite{CDI13}
for $so$ case. As we will see below
 this method works perfectly also for $sp$ case.

\vspace{.5cm}

{\it 2nd part: The symmetry condition and the generating function form}

The symmetry condition (\ref{sym}) implies that the spinorial $R$-
operator decomposes into the sum (\ref{Rrkosp-1}) over invariants which are tensor
products of the basis elements (\ref{basis}) of the algebra ${\cal A}$.
We write (\ref{Rrkosp-1}) in the concise form
\be \label{Rrkosp}
\check{\mathfrak{R}}
=\sum_
k\frac{r_k(u)}{k !} \sum_{\vec{a},\vec{b}} \; \e_{\vec{a},\vec{b}} \;\,
 c_1^{[a_1\ldots a_k)} \; c_2^{[b_1 \ldots b_k)} \; ,
\ee
where we have  denoted again the first and the
second factors of tensor the product by subscripts $1$ and $2$
and have introduced the shorthand notations
$$
\e_{\vec{a},\vec{b}} =\e_{a_1b_1}\cdots \e_{a_kb_k} \; , \;\;\;
c_1^{[a_1\ldots a_k)} = c_1^{[a_1}\cdots c_1^ {a_k)} \; , \;\;\;
c_2^{[a_1\ldots a_k)} = c_2^{[a_1}\cdots c_2^ {a_k)}   \; .
 $$


 After substituting (\ref{Rrkosp})
 the  Yangian-type
 condition (\ref{defR}) takes the form
\be\label{Rdef}
\begin{array}{c}
\displaystyle{
\sum_{k=0}^\infty\!\frac{r_k(u)}{k!} \sum_{\vec{a},\vec{b}}
 \e_{\vec{a},\vec{b}} \!\left(
u  \, \Bigl[ c_1^{[a_1 \ldots a_k)} \cdot c_2^{[b_1 \ldots
 b_k)}c_2^{[a b)}\!-\!c_1^{[a b)}c_1^{[a_1 \ldots a_k)} \cdot
 c_2^{[b_1 \ldots b_k)} \Bigr] \right.- }  \\ [0.3cm]
 \displaystyle{
 -\left.  \e_{fe} \, \Bigl[ c_1^{[a_1 \ldots a_k)}c_1^{[a e)} \cdot
 c_2^{[b_1 \ldots b_k)}c_2^{[f b)}
 -c_1^{[a e)}c_1^{[a_1 \ldots a_k)} \cdot
 c_2^{[f b)}c_2^{[b_1 \ldots b_k)} \Bigr] \!\right) = 0 \; . }
 \end{array}
 \ee

Using the representations (\ref{aab}),
(\ref{aba}) for the factors in the tensor products appearing in the
equation (\ref{Rdef}) we write this equation as
\be\label{Rdef1}
\begin{array}{c}
\displaystyle{
\sum_{k=0}^\infty\!\frac{r_k(u)}{k!} \;
\dd^k_\lambda \; e^{\lambda (\dd_2 \cdot  \dd_1)} \, \left(
u \,  \Bigl( [+]_2^{a b}
 \!-\! [-]_1^{a b} \Bigr) \right.   - }  \\ [0.3cm]
 \displaystyle{
 - \left.\left.  \e_{fe} \, \Bigl( [+]_1^{a e} \cdot
 [+]_2^{f b}  -[-]_1^{a e} \cdot [-]_2^{f b} \Bigr) \!\right)
 e^{(\kappa_1\cdot c_1)}  e^{(\kappa_2\cdot c_2)}
 \right|_{\lambda=\kappa_1=\kappa_2=0} = 0 \; , }
 \end{array}
 \ee
 where we have  applied also the  formula
 $$
\left.
 \dd^k_\lambda \; e^{\lambda (\dd_2 \cdot  \dd_1)} \right|_{\lambda=0} = (\varepsilon_{a b} \dd_2^b \,  \dd_1^a)^k =
\varepsilon_{a_1 b_1} \ldots \varepsilon_{a_k b_k}
\dd_1^{a_1 \ldots a_k} \; \dd_2^{b_1 \ldots b_k}  \; .
 $$

Our task is now to commute in (\ref{Rdef1}) all derivatives $\dd_i^a$
 to the right
 and then put to zero all variables $\kappa^a_i$
 appearing on left of the derivatives $\dd_i^a$.
 Taking into account (\ref{dkappa}) we have the rules
 \be
 \lb{nabla}
 e^{\lambda (\dd_2 \cdot  \dd_1)} \, \kappa_1^a =
 (\kappa_1^a + \lambda \, \dd_2^a) \, e^{\lambda (\dd_2 \cdot  \dd_1)} \; , \;\;\;
  e^{\lambda (\dd_2 \cdot  \dd_1)} \, \kappa_2^a =
 (\kappa_2^a + \epsilon \lambda \, \dd_1^a)
 \, e^{\lambda (\dd_2 \cdot  \dd_1)} \; ,
 \ee

\vspace{.5cm}

{\it 3rd part: Evaluating the Yangian-type condition}

 Applying these rules to the first term in (\ref{Rdef1}) we obtain
 \be\label{ke}
 \begin{array}{c}
 e^{\lambda (\dd_2 \cdot  \dd_1)} \,
  \Bigl( [+]_2^{a b} \!-\! [-]_1^{a b} \Bigr)   |_{\kappa=0}
 = \Bigl( \left[ \dd^a_2\dd^b_2
+ \lambda \frac12(  \dd^{a}_1 \dd^b_2 - \epsilon
\dd^{b}_1\dd^a_2 ) + \lambda^2 \frac14 \dd^{a}_1\dd^{b}_1\right] - \\ [0.2cm]
 - \left[ \dd^a_1\dd^b_1
- \lambda \frac12( \epsilon \dd^{a}_2\dd^b_1 -
\dd^{b}_2\dd^a_1 ) +\lambda^2 \frac14 \dd^{a}_2\dd^{b}_2\right] \Bigr)
 e^{\lambda (\dd_2 \cdot  \dd_1)}  =  \\ [0.2cm]
 =  (\lambda^2 \frac14 -1) \Bigl(\dd^{a}_1\dd^{b}_1 -  \dd^a_2\dd^b_2 \Bigr)
 e^{\lambda (\dd_2 \cdot  \dd_1)} \; ,
\end{array}
 \ee
 where after reordering we put to zero all variables $\kappa^a_i$
 appearing on left hand side with respect to derivatives $\partial_i^a$.

 The second term in (\ref{Rdef1}) contains the expression
$$
Y^{[ae)[fb)}(\kappa_1,\kappa_2) \equiv \Bigl( [+]_1^{a e} \cdot
 [+]_2^{f b}  -[-]_1^{a e} \cdot [-]_2^{f b} \Bigr) =
 $$
 $$
 = \left[ \dd^a_1\dd^e_1 +\frac14 \kappa^{a}_1\kappa^{e}_1 \right]
( \epsilon \kappa^{f}_2\dd^b_2 - \kappa^{b}_2\dd^f_2 ) +
( \epsilon \kappa^{a}_1\dd^e_1 - \kappa^{e}_1\dd^a_1 )
\left[ \dd^f_2\dd^b_2 +\frac14 \kappa^{f}_2\kappa^{b}_2 \right] \; ,
$$
 where we substituted (\ref{plumi}).
Further we act to this expression by the operator
$e^{\lambda (\dd_2 \cdot  \dd_1)}$ from the left and then
move $e^{\lambda (\dd_2 \cdot  \dd_1)}$ to the right with the help of
 rules (\ref{nabla}).
The result is
$$
\begin{array}{c}
 e^{\lambda (\dd_2 \cdot  \dd_1)}  \cdot Y^{[ae)[fb)}(\kappa_1^a,\kappa_2^a) =
 Y^{[ae)[fb)}(\nabla^a_2,\nabla^a_1)  \cdot e^{\lambda (\dd_2 \cdot  \dd_1)} \; ,
 \end{array}
$$
where
$\nabla^{a}_1 =  (\kappa_2^a + \epsilon \lambda \, \dd_1^a)$,
$\nabla^{a}_2 = (\kappa_1^a + \lambda \, \dd_2^a)$.
After reordering of variables $\kappa^a_i$ and $\dd^b_i$
 and cancelling all $\kappa_i^a$ appearing at the left we deduce
$$
\begin{array}{c}
\left. Y^{[ae)[fb)}(\nabla^a_2,\nabla^a_1) \right|_{\kappa_i = 0}
 \; =  \; \left[  \bigl( \epsilon \nabla_1^f \dd^b_2 -  \nabla_1^b \dd^f_2 \bigr)
 \dd^a_1\dd^e_1  +   \frac14 \nabla_2^a \nabla_2^e \,
 \bigl( \epsilon \nabla_1^f \dd^b_2 -  \nabla_1^b \dd^f_2 \bigr) +  \right. \\ [0.2cm]
 \left.\left.  + ( \epsilon \nabla_2^a \dd^e_1 -
 \nabla_2^e \dd^a_1 )
 \, \dd^f_2\dd^b_2 +\frac14 \nabla_1^f
 \nabla_1^b \; \bigl( \epsilon \nabla_2^a \dd^e_1 -
 \nabla_2^e \dd^a_1 \bigr) \right] \right|_{\kappa_i =0} = \\ [0.2cm]
 =   \lambda \left[ \dd^{ae}_1+   \frac14 \lambda^2 \dd_2^{ae} \right]
(\dd^{f}_1 \dd_2^b - \epsilon \, \dd^{b}_1 \dd_2^f) +
  \lambda ( \epsilon  \, \dd_2^a \dd^e_1 - \dd_2^e \dd^a_1 )
\left[ \dd^{fb}_2  +\frac14 \lambda^2 \, \dd_1^{fb}  \right]
+ \\ [0.2cm]
+ \frac14 \lambda^2 \Bigl( ( \varepsilon^{ef} \dd_2^{ab} +
 \varepsilon^{af} \, \dd_2^{be}
+ \varepsilon^{eb} \, \dd_2^{fa}
+  \varepsilon^{ab} \, \dd_2^{ef} ) -
( 2 \to 1  )  \Bigr) \; .
 \end{array}
 $$
 Here the formulas $\left. \dd^a_i \dd^e_i \kappa^f_i \right|_{\kappa_i =0} =
 \varepsilon^{fe} \dd_i^{a} - \varepsilon^{af} \dd_i^{e}$
 are helpful.

Note that we obtain the expression for
 $\left. Y^{[ae)[fb)}(\nabla^a_2,\nabla^a_1) \right|_{\kappa_i = 0}$
 which is (anti-)symmetric under the permutation of any three indices
 out of four $\{aefb\}$. This implies  that the same index symmetry holds for
 the
 operator $X^{[ae)[fb)}$ in (\ref{defR3}). This fact proves that the condition
 (\ref{defR5}) follows from (\ref{defR3}).

Finally, after contraction with $\varepsilon_{fe}$, we obtain
$$
\varepsilon_{fe} \left. Y^{[ae)[fb)}(\nabla^a_2,\nabla^a_1) \right|_{\kappa = 0} =
\left[ \frac{\lambda^2}{4} (n - 2\epsilon) -
\epsilon \lambda \Bigl(1 + \frac{\lambda^2}{4} \Bigr) (\dd_2 \cdot \dd_1)
\right]  (\dd_2^{ab} - \dd_1^{ab}) \; ,
$$
 and together with (\ref{ke}) we write the equation (\ref{Rdef1}) as
 following
 \be
 \lb{Rdef5}
  \begin{array}{c} \left.
 \sum\limits_{k=0}^\infty\!\frac{\eta_k \; r_k(u)}{k!} \;
 W_k((\dd_2 \cdot \dd_1)) \, (\dd_2^{ab} - \dd_1^{ab}) \,
 e^{(\kappa_1\cdot c_1 + \kappa_2\cdot c_2)}
 \right|_{\kappa_1=\kappa_2=0} = 0 \; ,
 \end{array}
 \ee
  where we have introduced the notation
  $W_k\bigl((\dd_2 \cdot \dd_1)\bigr)$ for the operator
  $$
  \left.
  W_k\bigl((\dd_2 \cdot \dd_1)\bigr) =
  \dd^k_\lambda  \, \left(
 u \,  \Bigl(1- \frac{\lambda^2}{4} \Bigr) -
 \frac{\lambda^2}{4} (n - 2\epsilon) +
 \epsilon \lambda \Bigl(1 + \frac{\lambda^2}{4} \Bigr)
 (\dd_2 \cdot \dd_1) \!\right) \; e^{\lambda (\dd_2 \cdot  \dd_1)}
 \right|_{\lambda=0} =
  $$
  $$
 \begin{array}{c}
\left.
= \dd^k_\lambda  \, \left(
 u  - \frac{\lambda^2}{4}
 (n - 2\epsilon + u) + \epsilon \lambda (1 + \frac{\lambda^2}{4}) \dd_\lambda
 \!\right)
 \; e^{\lambda (\dd_2 \cdot  \dd_1)}
 \right|_{\lambda= 0} =
 \end{array}
 $$
 \be
 \lb{Rdef6}
= \left.
\left( (u  + \epsilon k) \dd^k_\lambda  + \frac{k(k-1)}{4}
 (n - \epsilon k  + u)  \dd_\lambda^{k-2} \!\right)
 \; e^{\lambda (\dd_2 \cdot  \dd_1)}
 \right|_{\lambda= 0} \; .
 \ee
By substitution of the operator (\ref{Rdef6}) into
(\ref{Rdef5}) one obtains the recurrent relation
for the coefficients $r_k(u)$:
   \be
 \lb{Rdef7}
  r_{k+2}(u) = \frac{4 \,(k  + \epsilon \, u)}{((k+2)  - \epsilon (u+n))} \,
  r_k(u) \; .
 \ee
 The solution  of (\ref{Rdef7})  separates for even and odd $k$
 and it is now not hard to check that it is given by the formulas (\ref{Rdef8}).
 \hfill \qed

In the $so(n)$ case ($\epsilon = +1$) our results (e.g. relation (\ref{Rdef7}))
   coincide with the results of \cite{CDI13} after
   rescaling of generators $c^a \to \sqrt{2} c^a$
   which gives the standard definition of the Clifford
   algebra $c^a c^b + c^b c^a = 2 \, \epsilon^{ab}$.


We have encountered the additional condition (\ref{defR5}). Representations
for the generators of which it is fulfilled result in $L$ operators linear
in $u$ obeying the $RLL$ relation (\ref{rll-osc}) with the spinorial YB operator
${\mathfrak{R}}$ (\ref{Rrkosp-1}).

\noindent
 {\bf Proposition 6.} {\it
 The representations of Jordan-Schwinger type (\ref{jshm01}, \ref{jshm02})
$$
 M_{ab} = \epsilon(x_a \partial_b - \epsilon \, x_b \, \partial_a)=
   (\epsilon \, x_a \, \partial_b -x_b \, \partial_a) \; ,
 $$
obey the additional condition (\ref{defR5}), i.e.
$$  [M_{[ae} M_{f)b}]_+ = 0  \;. $$
Thus the $L$ operators
$$ L(u) = u I -F_{ab} M^{ab}, \tilde L(u) = u I + F^t_{ab} M^{ab} $$
built from the generators of the spinor representation $F_{ab}$ and of the
JS representation $M^{ab}$
obey the $RLL$ relation (\ref{rll-osc}) with the spinorial YB operator
${\mathfrak{R}}$ (\ref{Rrkosp-1}).
Here  $F_{ab}$ are understood as operators acting in the spinor space
and the superscript $t$, $F^t$, means  transposition.  }

\noindent
 {\bf Proof.}
The proof is done by straightforward calculations.
 \hfill \qed

\vspace{.5cm}

In Section 4 we have seen that the representations of JS type
obey the cubic characteristic identity for the matrix of its generators.
We see now that this identity and the additional condition (\ref{defR5})
are related.

\noindent
 {\bf Proposition 7.} {\it
If the generators $G^{a}_{\; b}$ of the algebra $so$, or $sp$,
 in a representation $\rho'$ obey the additional condition  (\ref{defR5}), i.e.
 \be
 \lb{defR5b}
   \rho' \Bigl( [G_{[ae} \, , \; G_{f)b}]_+ \Bigr) = 0  \; ,
 \ee
then the matrix $G = ||G^{a}_{\; b}||$
 (for simplicity here and below we omit the symbol
 of the representation $\rho'$)
 obeys the cubic characteristic identity
(cf. (\ref{char-3g}))}
 \be
 \label{cubic5}
G^3 + (\epsilon-n) G^2   + (\epsilon n-2) G
 + \frac{1}{2} {\rm Tr}(G^2) (I - \epsilon G) = 0  \; .
 \ee

 \vspace{.5cm}

\noindent
 {\bf Proof.}
It is convenient to rewrite (\ref{defR5b}) using the commutation
relation (\ref{sospg1}) as:
\be\label{cycl}
G_{a_2a_1}G_{c_1c_2}+G_{a_1c_1}G_{a_2c_2}+G_{c_1a_2}
G_{a_1c_2}=\e_{c_2a_1}G_{c_1a_2}+\e_{c_2a_2}G_{a_1c_1}
+\e_{c_2c_1}G_{a_2a_1}.
\ee
The multiplication
by $G^{a_1a_2}$ and summation over $a_1,a_2$ leads to
\be\label{cubic}
 \begin{array}{c}
m_2G_{c_1c_2} -2G_{c_1a_1}\e^{a_1a}G_{ab}\e^{ba_2}G_{a_2c_2}
 +2(n-2  \epsilon)G_{c_1b}\e^{ba_2}G_{a_2c_2} = \\ [0.2cm]
= -2G_{c_2b}\e^{ba_2}G_{a_2c_1}+m_2\e_{c_2c_1}  \; ,
 \end{array}
 \ee
where $m_2 = {\rm Tr} G^2$. From commutation
 relations (\ref{sospg1}) we also deduce
 $$
G_{c_2b}\e^{ba_2}G_{a_2c_1}=\epsilon[G_{c_1a}
\e^{ab}G_{bc_2}-(n-2\epsilon)G_{c_1c_2}],
 $$
 and applying this identity to the right hand side of (\ref{cubic})
  we obtain
  \be\label{cubic0}
 \begin{array}{c} G_{c_1a_1}\e^{a_1a}G_{ab}\e^{ba_2}G_{a_2c_2}- (n-2
\epsilon)G_{c_1b}\e^{ba_2}G_{a_2c_2}
 -\frac{1}{2} \, m_2G_{c_1c_2} = \\ [0.2cm]
 = \epsilon[G_{c_1a}
 \e^{ab}G_{bc_2}- (n-2\epsilon)G_{c_1c_2}
  -\frac{1}{2}m_2\e_{c_1c_2}],
 \end{array}
 \ee
or simply for the matrix $G = ||G^{a}_{\;\; b}|| =
 ||\varepsilon^{ac} \, G_{c b}||$
we have the characteristic relation
\be\label{cubic1}
G^3-(n-\epsilon) \, G^2+
 \epsilon \, \Bigl(n-2\epsilon-\frac{m_2}2 \Bigr) \, G
 +\epsilon \, \frac{m_2}2=0 \; ,
\ee
which coincides with (\ref{cubic5}). \hfill \qed

Note, that by parameterizing the eigenvalue of the quadratic
Casimir operator as in (\ref{specM2})
$$
m_2 \equiv {\rm{Tr}}G^2=2 \, m \, (m \epsilon +n-2 \epsilon),
$$
one can rewrite the polynomial (\ref{cubic5}) in the factorized form
\be\label{cubic2}
 \begin{array}{c}
G^3+(\epsilon - n)G^2
 +(\epsilon n-2-  \epsilon  \frac{m_2}2)G + \frac{m_2}2= \\ [0.2cm]
  =
  (G+ \epsilon m)(G -\epsilon m - n + 2\epsilon)(G - \epsilon)=0 \; .
 \end{array}
\ee
Note that this condition was presented in \cite{Re}
(see eq. (3.9) there)
for the case of $so(n)$ Lie algebra and for the generators
$G_{ab} \in so(n)$ which coincides
 with ours (satisfying the commutation relations
 (\ref{sospg1}) and (\ref{jshm03})) up to the redefinition  $G_{ab} \to -G_{ab}$.


An obvious generalisation of the Yangian concept discussed in Sect. 3 can be
considered where the fundamental $R$ matrix is replaced by the spinorial
$\mathfrak{R} $. The spinorial $L$ operators of the first order evaluation
of the spinorial Yangian $Y_s(\mathcal{G})$,
\be \label{LFG} L(u) = Iu - F_{ab} G^{ab}, \ee
where $F^a_b = \frac12 \e_{bd} c^{(a}c^{d]}$ are the spinor representation
generators and $G^{ab}$ are generators acting in $V$ and
obeying the condition (\ref{defR5}).
Monodromies built as products with multiplication in the spinor space $V_s$ but
tensor product of copies of $V$ result in examples of higher order
evaluations of the spinorial Yangian. Below we shall consider instead the
monodromy of two $L$ factors of this form but with the roles of the
representations $V_s$ and $V$ interchanged. The fusion procedure involving
the projection of $V_s\otimes V_s $ onto the fundamental representation
space $V_f$ results in examples of the second order evaluation of the
(fundamental) Yangian $Y(\mathcal{G})$.

	\section{Fusion operations}
	
Recalling the known procedure we consider the YB relation (\ref{YB})
with the representation by operators in the space $V_1 \otimes V_2 \otimes
V_3 $. If it holds for the particular choices for $R_{i,3}$ as $L_{i,3}$
and $L_{i, \tilde 3}$ then it will hold also for the choice
\be \label{Ti33}
 T_{i, 3 \tilde 3} (u) = L_{i,3}(u-\lambda) L_{i, \tilde 3}(u-\mu)
\ee
defined by the product of operators in the space $V_i$ and as tensor product
action in $V_3 \otimes V_{\tilde 3}$ . Consider then the projection $\Pi$ on the
invariant subspace $V_{\Pi 3}=\Pi \cdot V_3 \otimes V_{\tilde 3}$ and the restriction
of the operator $T_{i, 3 \tilde 3} (u)$ to this subspace $L_{i,\Pi 3}$.
$$
((V_3 \otimes V_{\tilde 3}) \otimes (V_3 \otimes V_{\tilde 3})^t
\to V_{\Pi3}\otimes V_{\Pi3}^t.
$$
The
YB relation then holds for the substitution of $R_{i,3}$ by $L_{i,\Pi 3}$.

We consider first the case with the fundamental representation for both
$V_1$ and $V_2$. $V_3$ and $V_{\tilde 3}$ are both the spinorial spaces and
$$ L_{i,3}(u) = u I - F, \ \ L_{i,\tilde 3}(u) = u I + F^t $$
The fermionic oscillators used above to express the spinorial generators
are related to the conventional gamma matrices and the transposition $t$ is
defined in the matrix sense.
\be \label{Fgamma}
F^a_b = \frac12 \e_{bd} c^{(a}c^{d]}= \frac14 \e_{bd} \gamma^{a d},
\gamma^{ab} = \gamma^{(a}\gamma^{d]}, \gamma^a = \sqrt 2 c^a \ee

The gamma matrices represent the intertwiner of the fundamental representation
space $V_f$
labelled by the index $a$ and the corresponding invariant subspace in the tensor
product of the spinor spaces $V_s \otimes V_s$ labelled by two matric
indices, i.e. they project $V_s \otimes V_s \to V_f$. For the reduction of
the product of $L$ matrices the projection
$
((V_s \otimes V_s) \otimes (V_s \otimes V_s)^t \to V_f\otimes V_f^t
$
by contracting the spinor indices with
$  \gamma^{a_2 \alpha_2}_{\alpha_1} \gamma^{\beta_1}_{b_1 \beta_2}
$
is to be done.

This operation is performed by calculating gamma matrix traces of products
with 2, 4 and 6 $\gamma$, e.g.
$$ {\rm{tr}}(\gamma^a \gamma^b) = C\  \varepsilon^{ab}
{\rm{tr}} (\gamma^a \gamma^b \gamma^c \gamma^d)   = C\ \left (\varepsilon^{ab}
\varepsilon^{cd} -  \varepsilon^{ac} \varepsilon^{bd} + \varepsilon^{ad}
\varepsilon^{bc}\right ) $$
which follows from the Clifford anti-commutation relation with
$C$ being the trace of the unit matrix in spinor space.

The symplectic counterpart of the spinor representation is
infinite-dimensional. Nevertheless, the operators
$\Gamma_a $ can be interpreted as projectors of the tensor product
to the fundamental representation labelled by $a$. The matrix elements can be
calculated with respect to the standard basis of oscillator states or in
coherent states.
The infinite sum or integration over the basis states
requires a regularisation.
Here we  need only that this regularisation can be done in such a way
that the analogous relations for the   traces hold.
The following result relies on the relations
$$ {\rm{tr}} (\gamma^a \gamma^b) = C \varepsilon^{ab},
{\rm{tr}} (\gamma^a \gamma^b \gamma^c \gamma^d)   = C
\left (\varepsilon^{ab}
\varepsilon^{cd} - \e \varepsilon^{ac} \varepsilon^{bd} + \varepsilon^{ad}
\varepsilon^{bc} \right ) $$


\noindent
 {\bf Proposition 8.}
{\it
The fundamental $R$ matrix with $so$ or $sp$ symmetry (\ref{rzW},\ref{rzW1})
is reproduced by the fusion procedure applied to the product
$L(u-\mu) \tilde L(u-\lambda)$ of
the spinor $L$ matrix (\ref{L-1}) and its transposition
involving the projection of the tensor product of the two spinor
representation spaces to the invariant subspace corresponding to the
fundamental (spin 1 in $so$ case) representation. }

\vspace{.5cm}

\noindent
 {\bf Proof.}
Let us write the monodromy and the projection with explicit
indices for the $so$ case. Indices $a,b, ..$ label the basis of the
fundamental representation space and $\a, \b ...$ the basis of the spinor
spaces.
$$ (L_{1,f}(u))^{a_1 a_2}_{b_1 b_2}  = L_{1,3 \ c_1 \beta_1}^{a_1 \alpha_1}(u-\lambda)
L_{1, \tilde 3 b_1 \alpha_2}^{c \beta_2}(u-\mu) \times
\gamma^{a_2 \alpha_2}_{\alpha_1} \gamma^{\beta_1}_{b_1 \beta_2}
$$
We obtain
$$	(u-\lambda)(u-\mu)\delta^{b_1}_{d_1}{\rm{tr}}(\gamma^{b_2}\gamma_{d
		_2})-\frac12(u-\mu){\rm{tr}}(\gamma^{b_2}\gamma^{b_1}{}_{d_1}\gamma
	_{d_2})+\frac12(u-\lambda){\rm{tr}}(\gamma^{b_1}{}_{d_1}\gamma^{b_2}
	\gamma_{d_2})-
	$$
	$$
	-\frac14{\rm{tr}}(\gamma^{c_1}{}_{d_1}\gamma^{b_2}\gamma^{b_1}{}_{c
		_1}\gamma_{d_2}) = C \cdot
$$ $$
[\{(u-\lambda)(u-\mu)-\frac{n-3}4 \}\delta
^{b_1}_{d_1}\delta^{b_2}_{d_2}+\{u-\frac{\lambda+\mu}2+\frac{n-2}4\}
\delta^{b_1}_{d_2}\delta^{b_2}_{d_1}-\{u-\frac{\lambda+\mu}2-\frac
{n-2}4\}\delta^{b_1b_2}\delta_{d_1d_2}].
$$
In the symplectic case we have the analogous calculation and the result for
both cases is proportional to
$$
\{(u-\lambda)(u-\mu)-\frac{\e n-3}4\}
I +\{u-\frac{\lambda+\mu}2+\frac{\e n-2}4\}
P -\{u-\frac{\lambda+\mu}2-\frac
{\e n-2}4\}\e K.
$$
If we impose
$
\lambda+\mu=\frac12(2-\e n)
$
and
$
\lambda\mu=\frac14(\e n-3)
$
we  recognize the fundamental $R$-matrix (\ref{rz}).
\hfill \qed

\vspace{1cm}

\noindent
 {\bf Proposition 9.} {\it
 Consider the product $L(u-\mu) \tilde
L(u-\lambda)$ of the spinorial $L$ operators (\ref{LFG})
$$ L(u) = u I -F_{ab} G^{ab}, \tilde L(u) = u I + F^t_{ab} G^{ab} $$
where $F^a_b $ are the spinor representation
generators and $G^{ab}$ are generators
obeying the condition (\ref{defR5}), i.e.
$$  [G_{[ae} G_{f)b}]_+ = 0  \; $$
The application of the fusion procedure
involving the projection of the tensor product of the two spinor
representation spaces to the invariant subspace corresponding to the
fundamental  representation results in
the $L$ operator of the form of the quadratic evaluation type,
  which at the shift parameter values
 \be\label{lm}
(\lambda-\mu)^2=\frac{(n-4)^2}4.
\ \
\lambda+\mu=0,
 \ee
 reads
\be \label{LGN}
L(u)=u^2 I+u G +N,
 \ee
$$ N=\frac12 (G^2-\beta G) -\frac14 \beta^2 I -\frac{m_2}8,  \ \
\beta = \frac{n}{2} - \e.
$$ }

\vspace{.5cm}

\noindent
 {\bf Proof.}
We may start from the YB relation with the JS representation in
$V_1$, and the spinorial in $V_2, V_3$ or from the $RLL$ relation with
the JS representation in $V_1, V_2$ and the spinorial one in $V_3$. In the
first case
the fusion operation will lead in
particular to $T_{1,3\tilde 3}$ (\ref{Ti33})
with JS operator   multiplication for the action in $V_1$ and  in the tensor
product of two copies of spinorial representations $V_3$.
The projection of the latter to the fundamental representation
is done by just the same calculation as above.

Let us write the monodromy $T_{1,3\tilde 3} $  and the projection with explicit
indices for the $so$ case.
$$
[(u-\lambda){\mathbb{I}}-\frac14\c^{ab}G_{ab}]^\a_\b[(u-\mu){
\mathbb{I}}+\frac14\c^{cd}G_{cd}]^\c_\d .
$$
 The contraction with $(\c^e)^\b_\c(\c^f)
^\d_\a$ leads to
$$
{\rm{L}}(u)=(u-\lambda)(u-\mu){\rm{tr}}(\c^e\c^f)-\frac14(u-\mu){\rm
{tr}}(\c^{ab}\c^e\c^f)G_{ab}+\frac14(u-\lambda){\rm{tr}}(\c^e\c^{cd}
\c^f)G_{cd}-
$$
$$
-\frac1{16}{\rm{tr}}(\c^{ab}\c^e\c^{cd}\c^f)G_{ab}G_{cd}={\rm{tr}}{
\mathbb{I}}\cdot[(u-\lambda)(u-\mu)\d^{ef}+\frac12(u-\mu)G^{ef}+
\frac12(u-\lambda)G^{ef}+
$$
$$
+\frac14(G^{eb}G^{bf}+G^{fb}G^{be}-\frac12\d^{ef}G^{cd}G_{dc})].
$$
By a shift $u \to u-\lambda$ this can be written as
$$ u^2\d^{ef}+u(G^{ef}-(\mu+\lambda)\d^{ef})+N^{ef},
$$
 \be\label{nef}
N^{ef}=\frac12(F^2)^{ef}-\frac18\d^{ef}{\rm{tr}}G^2+(\frac{n-2}4-
\frac{\mu+\lambda}2)G^{ef},
 \ee
Here we have used the Lie algebra relations to transform the commutator of generators
$G$.
\hfill \qed

\vspace{.5cm}

By direct calculation it is checked that the
obtained $L$ operator at the particular parameter values
(\ref{LGN})
obeys the $RLL$ relation (\ref{rll})
with the fundamental $R$ matrix (\ref{rzW1}).

\section{Summary  }
\setcounter{equation}{0}

We have considered Yang-Baxter $R$ operators symmetric with respect to the
orthogonal and symplectic algebras. We have started from known examples
illustrating how the more involved structure of these algebras compared to
the one of the special linear type result in more involved features of the
$R$ operators. We have shown how both cases can be treated in a uniform way, which
amounts in particular in the interchange of symmetrisation with
anti-symmetrisation. It is known that this feature of analogy allows a
supersymmetric formulation starting from the graded orthosymplectic algebra.
We have preferred the more explicit parallel treatment of the two cases and
decided not to add the supersymmetric formulation here.

The $L$ operators obeying the $RLL$ relation together with the $so$ or $sp$
symmetric fundamental $R$ matrix define the corresponding Yangian algebra.
Unlike the case of $s\ell$ symmetry the truncation of the expansion of
$L(u)$ in inverse powers of the spectral parameter $u$ results in
constraints, which cannot be fulfilled in the enveloping algebra, but lead
to the
 restriction  to distinguished  representations the generators of which can build  such
 $L$ operators.

The known example of truncation at the first order, the linear evaluation of
the Yangian, is given by the spinor representation of the orthogonal algebra.
It can be formulated on the basis of a fermionic oscillator or Clifford
algebra. We have indicated its symplectic counterpart
(metaplectic representation), which is formulated
on the basis of a bosonic oscillator algebra. The constraint resulting form
the
first order truncation can be formulated as
 a characteristic identity of second order in terms of the matrix of
generators.

The fundamental $R$ matrix can be regarded as an example for the trucation at
the second order, the quadratic evaluation of the Yangian algebra.  The
Jordan-Schwinger type representations provide more examples.
The constraint resulting form the
second  order truncation can be formulated as
 a characteristic identity of third order in terms of the matrix of
generators.

The YB relation involving the spinor and
metaplectic representation $L$ operators together
with the particular $R$ operator acting in the tensor product of two spinor and metaplectic
representations has been studied. On its basis the explicit form of this
spinorial and metaplectic
 $R$ operator has been derived in an uniform treatment of both
the orthogonal and symplectic  cases. Further, we have studied
a similar YB relation involving this spinorial $R$ operator together with YB operators
acting in a tensor product space of the  the spinor
(and metaplectic) with  another representation
different from the fundamental one. The demanded YB relation results in a
constraint on the generators of this representation. It is fulfilled by the
fundamental representation, but also by Jordan-Schwinger type
representations. For the constraint the latter have to be based on
bosonic Heisenberg algebras
in the orthogonal case and on fermionic Heisenberg algebras in the
symplectic case.
We have shown that the latter constraint is directly related to the
third order characteristic identity of the quadratic Yangian evaluation.

We have studied fusion operations on products of YB operators acting on
tensor products where one tensor factor is the spinor representation and the
fusion involves the
 the projection of the tensor product of two spinor
respresensations onto the fundamental (vector) representation.
In particular we have demonstrated how the fundamental $R$ matrix is
reproduced performing the fusion operation of the product of spinor
$L$ operators. The fusion operation with the same projection has been
done also on the product of the $R$ operators obeying the above YB relation
with the spinorial $R$ operator. These examples of fusion are choosen,
because they result in the examples of $L$ operators of the quadratic
Yangian evaluation considered before, the fundamental $R$ matrix
and the $L$ operator of the
Jordan-Schwinger type.  The explicit form of the latter is found in this
way.

Yang-Baxter operators and in particular $L$ operators of simple structure,
which can be formulated explicitly, are of interest for integrable quantum
systems. In particular the monodromy operators defined as products of $L$
operators are applied in the investigation of integrable interaction  models
and in the construction of symmetric correlators and operators.

\vspace{0.5cm}

\noindent
{\bf Acknowledgments.}
The work of A.P.I. was supported by
RFBR grants 14-01-0047-a and  15-52-05022 Arm-a.
The work of D.K. was partially supported by the Armenian
State Committee of Science grant SCS 15RF-039. It was done
within programs of the ICTP Network NET68 and of the
Regional Training Network on Theoretical Physics
sponsored by Volkswagenstiftung Contract nr. 86 260.
 Our collaboration has also been supported by JINR (Dubna) via a
Heisenberg-Landau grant (A.P.I. and R.K.) and a
Smorodinski-Ter-Antonyan grant (A.P.I. and D.K.).

\end{document}